\documentclass[namedreferences]{SolarPhysics}
\usepackage[optionalrh]{spr-sola-addons}
\usepackage{courier}
\usepackage{graphicx}
\usepackage{natbib}
\usepackage{amssymb}
\usepackage{url}
\usepackage[usenames]{color}
\begin{document}

\begin{article}

\begin{opening}

\title{Is null-point reconnection important for solar flux emergence?}
%Magnetic Null Point Evolution in a Modelled Solar Flux Emergence Event

\author{R.~C.~Maclean$^{1}$\sep
        C.~E.~Parnell$^{1}$\sep
        K.~Galsgaard$^{2}$}

\runningauthor{Maclean et al.}
\runningtitle{Null-Point Reconnection in Flux Emergence}

\institute{$^{1}$ \mbox{Institute of Mathematics,} \mbox{University of St Andrews,} \mbox{The North Haugh,} \mbox{St Andrews,} Fife, \mbox{KY16 8YL}\\
email: \url{rhonam@mcs.st-andrews.ac.uk}\\
$^{2}$ \mbox{Niels Bohr Institute,} \mbox{Julie Maries vej 30,} \mbox{2100 Copenhagen,} Denmark
}

\begin{abstract}
The role of null-point reconnection in a 3D numerical MHD model of solar emerging flux is investigated. The model consists of a twisted magnetic flux tube rising through a stratified convection zone and atmosphere to interact and reconnect with a horizontal overlying magnetic field in the atmosphere. Null points appear as the reconnection begins and persist throughout the rest of the emergence, where they can be found mostly in the model photosphere and transition region, forming two loose clusters on either side of the emerging flux tube. Up to 26 nulls are present at any one time, and tracking in time shows that there is a total of 305 overall, despite the initial simplicity of the magnetic field configuration. We find evidence for the reality of the nulls in terms of their methods of creation and destruction, their balance of signs, their long lifetimes, and their geometrical stability. We then show that due to the low parallel electric fields associated with the nulls, null-point reconnection is not the main type of magnetic reconnection involved in the interaction of the newly emerged flux with the overlying field. However, the large number of nulls implies that the topological structure of the magnetic field must be very complex and the importance of reconnection along separators or separatrix surfaces for flux emergence cannot be ruled out.
\end{abstract}


\end{opening}

\section{Introduction}
\label{sec:intro}

The continual injection of new magnetic flux, energy, and helicity into the Sun's atmosphere from the convection zone below is fundamentally responsible for much of the activity observed in the solar atmosphere. As the newly emerged flux interacts with the pre-existing magnetic field above, X-ray jets, bright points, or active-region loop systems may be formed (depending on the spatial scale of the emergence), thus these features are a result of magnetic reconnection and so are the realisation of the heated corona.

\subsection{MHD modelling of flux emergence}

Numerical simulations are a powerful tool for modelling and understanding such complex inherently 3D systems. There is a large body of literature using magnetohydrodynamic (MHD) codes to model flux emergence events. Magnetic flux is created by the solar dynamo in the convective overshoot layer at the base of the convection zone, known as the tachocline \citep{1992A&A...265..106S}. Instabilities cause tubes of flux to become buoyant and rise through the convection zone towards the solar surface. During the initial rise phase, the thin-flux-tube approximation has been used to predict active-region emergence latitudes and tilt angles that are consistent with observations \citep{1994A&A...281L..69S,1995ApJ...441..886C,2006MNRAS.369.1703H}. However, as the flux tubes approach the photosphere they expand significantly, so the thin-flux-tube approximation is no longer valid, as the flux tubes are likely to become fragmented or at least severely distorted, and full 3D MHD models are required. The magnetic fieldlines must have enough twist to prevent the interactions with the surrounding medium from fragmenting the flux tube \citep{1996ApJ...464..999L}, although the critical degree of twist required depends on the apex curvature of the loop and can be significantly less in 3D than the 2D limit implies \citep{2000ApJ...540..548A}.

The emergence in three dimensions of a single flux tube or sheet into an initially field-free atmosphere was first studied by \citet{1992PASJ...44..167M}, who obtained many basic features of the emerging field, such as draining of plasma down the fieldlines, expansion of loops into the corona, and formation of shock waves at the loop footpoints. Later work on this topic by several authors, all using a stably stratified convection zone, a low temperature photosphere and a high temperature (but field-free) corona, went into greater detail \citep[see ][]{2005ApJ...635.1299A}. \citet{2003ApJ...582..475A}, using an anelastic MHD model of the convection zone coupled to a fully compressible code for the atmosphere, found that the newly-emerged coronal fieldlines formed sigmoidal structures whose chirality depended on where in the atmosphere the emitting plasma was assumed to be located. \citet{2003ApJ...586..630M} showed that the initially bipolar photospheric magnetic field structure can develop into a quadrupolar structure, and developed the classification of emerged fieldlines as either expanding or undulating. Finally, \citet{2004ApJ...610..588M} found that the natural shearing motions that occur as the field expands into a pressure-stratified atmosphere tend to add axial flux and decrease the twist of the fieldlines.

However, the real corona contains a pre-existing magnetic field, and interesting results have been obtained by including an overlying field in flux emergence models. In 2.5D, \citet{1996PASJ...48..353Y} experimented with vertical, oblique, and several orientations of horizontal background coronal magnetic fields. All the configurations lead to a current sheet forming over the emerging loops, and the production of magnetic islands and jets via the tearing instability. However, as \citet{2003ApJ...586..630M} have pointed out, full 3D models are required to adequately reproduce the draining of plasma down the magnetic fieldlines. Such simulations have been carried out by, for example, \citet{2004A&A...426.1047A,2005ApJ...635.1299A}, who allowed a twisted flux tube to emerge into a uniform horizontal background magnetic field. They found that in such a 3D setup, magnetic reconnection occurred at multiple sites along fieldlines, rather than at a single isolated magnetic null as is the norm in 2D. This model reproduced fan-like arcades similar to those observed by TRACE, and also reconnection jets whose temperatures and velocities were a good fit with observations.

\citet{2004ApJ...609.1123F} also emerged a 3D twisted magnetic flux tube into a pre-existing coronal field, although here the coronal field was a potential arcade. Their model does not include a convection zone; the flux tube is `injected' into the atmosphere by controlling the value of the electric field ($\mathbf{v} \times \mathbf{B}$) at the lower boundary, thus specifying the electric current and hence the Lorentz force. This allows them to achieve full emergence of the tube into the atmosphere, by analytically defining the shape of the field, and the velocity with which it passes through the boundary. Although the ideal MHD equations were used, magnetic reconnection could still take place at current sheets due to numerical diffusion in the presence of strong gradients in the magnetic field. Once a certain critical amount of twist had emerged, the flux tube underwent the kink instability and began to accelerate upwards and twist around, producing a sigmoidal current sheet beneath. This evolution was found to be strongly dependent on the relative orientations of the emerging flux tube and the pre-existing coronal magnetic field: if the overlying field was oppositely-directed to the axial component of the flux tube, reconnection started straightaway and destroyed the flux tube as it rose into the atmosphere. The effect of the orientation of the overlying field was investigated by \citet{2007ApJ...666..516G}, using the same model as the Archontis papers above, but with different orientations of the plane-parallel coronal field relative to the emerging flux tube. They found that most reconnection took place when the two flux systems were close to anti-parallel, and hardly any when they were close to parallel, but, in all cases, the height reached by the emerging flux tube, as a function of time, was unchanged.

The models described so far all made certain simplifications about the behaviour of the convection zone, since their main aim was to investigate the effects of emergence into the solar atmosphere. However, the most realistic models currently possible of magnetic flux rising through the convection zone include radiative transfer, convection, compressibility, and thermal conduction. For example, \citet{2007A&A...467..703C} showed how a single initial rising flux tube influenced by convection leads to the emergence of many small flux bundles, which produce a characteristic (observed) dark lane appearing in the photospheric granulation pattern. However, since they have no atmosphere above the photosphere they could not study the effects of emergence on the atmosphere itself. \citet{2008ApJ...679..871M} recently performed a simulation in which their corona was heated by the release of stress built up by convective motions in a complex magnetic field topology. The granulation cells became enlarged as an emerging flux tube passed through them, in agreement with observations.

Recent work by \citet{2007A&A...470..709M} has shown that in fact such a high level of complexity in the subsurface field is not necessary for emergence models to give valid results in the solar atmosphere. They create a complex subsurface field by allowing two magnetic flux tubes to interact in a stably stratified convection zone before emerging, and show that the resulting emerged atmospheric magnetic field is very similar to the result for a simple single flux tube. This means that the results obtained from simpler models with stably-stratified convection zones are adequate when the aim is to study the resulting atmospheric magnetic fields.

\subsection{The next step: magnetic topology}

As we have just seen, previous work on solar emerging flux has greatly improved our understanding of many of the complex processes involved. But there is one key issue that was not addressed in all the previous analyses; they do not account for the topology of the magnetic fields. Here, by magnetic topology, we mean specifically the location and evolution of topological features such as magnetic null points, spines, separatrix surfaces, and separators \citep[see][for a good review]{2005LRSP....2....7L}. Plotting individual fieldlines that are not part of the topological skeleton, even if they are carefully selected, is not enough to ensure that all the information about the structure of the magnetic field, and the way it connects, is known.

Magnetic reconnection preferentially occurs at topological features including null points \citep{2004GApFD..98..407P,2005GApFD..99...77P,2007JGRA..11203103P}, separatrix surfaces \citep{2002ApJ...576..533P,2005ApJ...624.1057P}, and separator fieldlines \citep{2005ApJ...624.1057P,2007RSPSA.463.1097H,2008ApJ...675.1656P}, as well as at their geometrical counterparts: quasi-separatrix layers \citep{2006SoPh..238..347A,2006AdSpR..37.1269D,2007Sci...318.1588A,2007ApJ...660..863T} and quasi-separators \citep[also known as hyperbolic flux tubes;][]{2003ApJ...582.1172T,2003ApJ...595..506G,2005A&A...444..961A,2006A&A...451.1101D,2006A&A...459..627D,2007A&A...473..615W}. This is because the hyperbolic magnetic topologies around such features tend to focus the electric current, and a strong electric field (parallel to the magnetic field) is associated with 3D magnetic reconnection. So knowing the topological structure of the field provides a good guide to the probable locations of reconnection sites in the system.

In this paper, we concentrate on determining the importance of reconnection at magnetic null points in our model magnetic field. The signature of 3D magnetic reconnection is the existence of a strong electric field parallel to the magnetic field \citep{1988JGR....93.5547S}. We therefore study the parallel electric field near the nulls, to determine if reconnection takes place there. We also consider the current density, which may be parallel ($\mathbf{j}_{\parallel}$) or perpendicular ($\mathbf{j}_{\perp}$) to the spine of the null point, or a combination of both, and the consequences of this for the structure of the surrounding fieldlines are discussed in Section~\ref{ssec:locnat}. 3D magnetic reconnection may take place in the presence or absence of magnetic null points \citep{1988JGR....93.5547S,1988JGR....93.5559H,1995JGR...10023443P,1996JGR...101.7631D,1996RSPSA.354.2951P,2003PhPl...10.2717H,2006RSPSA.462.2877W}. However, in the flux emergence model that we study here, the change of magnetic topological structure is a central feature, and is cospatial with the location of magnetic reconnection. To have a changing magnetic topology, nulls are often involved, and hence we focus on reconnection at nulls in this paper. Reconnection at other topological features in our flux emergence experiment may also be significant, and this will form the subject matter for a future investigation.

So, the magnetic topology is important as it allows us to identify probable sites of magnetic reconnection. However, it has only been realised very recently that information about the topological structure is also required to make sense of the reconnection rate and energy release sites. Knowing the rate at which magnetic reconnection takes place is crucial to our understanding of how the magnetic field evolves, and recent work by \citet{2008ApJ...675.1656P} has shown that knowledge of the magnetic topology is absolutely essential for a correct interpretation of the results of the model. This is because, in a complicated magnetic field containing fieldlines with many different connectivities, a phenomenon called \emph{recursive reconnection} can take place. Recursive reconnection means that the same magnetic flux can be recycled many times through a repeating sequence of different magnetic connectivities. In a situation like this, knowledge of the topological structure of the magnetic field is vital to determine the global reconnection rate, which can in fact be much higher than would be determined via other methods.

In this paper, we take the first steps towards a full topological analysis of a flux emergence model. The dataset that we use is one of the numerical MHD models of \citet{2007ApJ...666..516G}, which consists of a twisted buoyant magnetic flux tube rising through the upper layers of a stably-stratified convection zone and into an atmosphere with a horizontal plane-parallel pre-existing magnetic field. Now, all of the fieldlines in the topological skeleton of the magnetic field must either start or end at magnetic null points (there are no bald patches \citep{1993A&A...276..564T} in our model as no fieldlines leave the closed boundaries of our box, only the periodic boundaries). We locate the null points in the magnetic field using the accurate new algorithm described by \citet{2007PhPl...14h2107H}. This paper concentrates on the surprising nature of these magnetic null points: their number, type, distribution, and evolution. Section~\ref{sec:model} describes the code, the model setup, and how it was analysed. Our results are reported in Section~\ref{sec:results} and then we conclude with a discussion in Section~\ref{sec:conclusions}.

\section{Model setup and analysis techniques}
\label{sec:model}

\subsection{Flux emergence model}
\label{ssec:femodel}

The specific flux emergence model that we analyse in this paper has already been described by \citet{2007ApJ...666..516G}, with more details given by \citet{2005ApJ...618L.153G,2005ApJ...635.1299A}. The latter two papers describe a slightly different version of the experiment, with a different interaction angle between the flux tube and the overlying field. We will summarise the most important features here. The background medium is made up of four horizontal layers; an adiabatically-stratified convection zone, a relatively cool isothermal photosphere, a transition region with a steep temperature gradient, and a hot isothermal corona. The initial magnetic field configuration consists of a horizontal overlying magnetic field, above the lower transition region, and a twisted magnetic flux tube which is inserted about $2\,$Mm below the base of the photosphere. This flux tube is horizontal in orientation, with a uniform twist around the tube axis, and a longitudinal component with a Gaussian profile. The chosen twist is such that the azimuthal component of the tube's magnetic field in the $(x,z)$-plane is given by $B_{\phi} = \alpha r B_x$, where $B_x$ is the longitudinal field, and $\alpha = 0.4$. It is stable to the kink instability, and its rise is triggered by the introduction of a density deficit which reaches its maximum at the central point and falls off as a Gaussian in both directions along the tube axis. As the flux tube rises, it eventually interacts with the horizontal overlying field, which is oriented so that the two flux systems come into contact with an angle of approximately $135^{\circ}$ between their leading fieldlines. This is a generic case \citep[case B from][]{2007ApJ...666..516G}, which allows for significant magnetic reconnection to take place between the two flux systems, but not the maximum possible amount, which takes place when the contact angle is $180^{\circ}$ \citep[as shown in ][]{2007ApJ...666..516G}.

\begin{figure}
  \centering
  \includegraphics[width=0.7\textwidth]{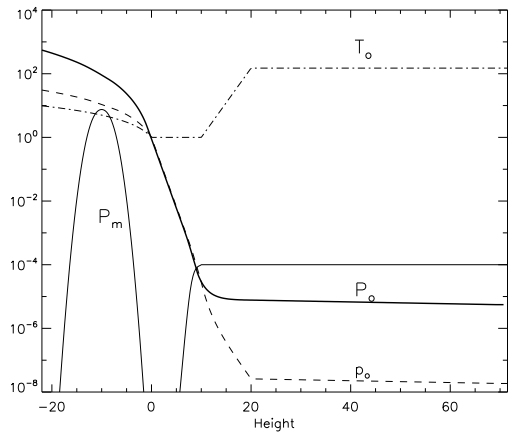}
  \caption{Initial values of the properties of the system along the central vertical line; gas pressure (thick solid line), magnetic pressure (thin solid line), temperature (dot-dashed line), and density (dashed line). Reproduced with kind permission from \protect\citet{2007ApJ...666..516G}.}
  \label{fig:pressure}
\end{figure}

The initial setup for the model can easily be visualised with reference to Figure~\ref{fig:pressure}. Throughout the paper, we use units that have been normalised according to the values in the photosphere: $p_{ph} = 1.4 \times 10^5 \,$ergs cm$^{-3}$, $\rho _{ph} = 3 \times 10^7 \,$g cm$^{-3}$, $T_{ph} = 5.6 \times 10^3 \,$K, and $H_{ph} = 170 \,$km. Other units used here include time ($t_{ph} = 25 \,$s), velocity ($V = 6.8 \,$km s$^{-1}$), and magnetic field ($B_{ph} = 1.3 \times 10^3 \,$G).

The system is set up on a numerical grid with (160, 148, 218) points in the $(x, y, z)$ directions respectively. Each cell in the $x$-direction corresponds to $148.8 \,$km, and for the $y$-direction this figure is $137.8 \,$km. The grid is stretched in the $z$-direction, so that the highest grid resolution in this direction (corresponding to $47.7 \,$km per grid cell) is concentrated in the region spanning the top of the convection zone to the base of the corona; the resolution is lower near the boundaries of the box. These cell sizes mean that the whole box is $23.8 \,$Mm $\times 20.4 \,$Mm $\times 15.6 \,$Mm, which in our simulation units gives a numerical domain covering $(-70, 70)$ in $x$, $(-60, 60)$ in $y$, and $(-22, 70)$ in $z$. The side boundaries of the box are periodic, and the top and bottom boundaries are closed.

The variables are evolved in time by solving the 3D, time-dependent, resistive MHD equations, using the code developed by \citet{Nordlund+Galsgaard95mhd}. The equations are solved numerically using high-order finite differencing; sixth-order accurate spatial derivatives are calculated using data from six neighbouring grid points, and corresponding fifth-order accurate interpolation routines are used. A third-order predictor-corrector algorithm advances the solution in time. The high spatial order in the code means that steep gradients in any of the variables can lead to numerical ringing and overshooting, so spatially-localised artificial viscosity and magnetic resistivity, along with a discontinuous shock-capturing mechanism, are introduced to prevent this. (In Section~\ref{sssec:stability} and Figure~\ref{fig:bxyzvar} we demonstrate how well the artificial viscosity and resistivity remove spurious short length scales.)

The evolution of the magnetic flux tube as it rises through the atmosphere and interacts with the coronal magnetic field has been described in detail by \citet{2007ApJ...666..516G}. As the flux tube rises and begins to interact with the overlying magnetic field, an archlike current sheet forms between the two flux systems. High-velocity outflows travel out from the current sheet along the direction of the overlying magnetic field. The flux tube apex rises slowly at first, then goes through a period of rising more quickly for a while, and finally slows down again towards the end of the experiment. However, the centre of the flux tube never fully emerges into the atmosphere, becoming stuck somewhere in the photospheric layer. By the end of the experiment, 65\% of the initial magnetic flux in the flux tube has emerged into the corona, and of this, 60\% has undergone magnetic reconnection and changed its connectivity.

\subsection{Determining the location and nature of the magnetic null points}
\label{ssec:locnat}

The magnetic null points present at each time step of the model were located using the trilinear algorithm presented by \citet{2007PhPl...14h2107H}, which has been used on observational data by \citet{2008A&A...484L..47R,2009SoPh..254...51L,cmn09}. Locating the null points proceeds in three stages. In the reduction stage, any grid cell that has the same sign for any one of the three components of the magnetic field at all eight cell corners is ruled out from further consideration as it cannot contain a magnetic null point. Then in the analysis stage, the existence of a null in each selected cell is confirmed or disproved by considering the intersection curves inside each cell where two components of the magnetic field are zero --- say $B_x = B_y = 0$. A null point must exist somewhere along such a curve in a grid cell if the third magnetic field component, in this case $B_z$, is of opposite sign at the two endpoints where the curve exits the grid cell. Finally in the positioning stage, the exact location of each detected null is pinned down using an iterative Newton-Raphson method.

The trilinear method of null-finding has significant advantages over the previously popular method of \citet{Greene1992194}, which uses the Poincar{\'e} index, and has been shown to arbitrarily create and destroy null points and move them several grid cells away from their true position in moderately nonlinear magnetic fields where the trilinear method is still accurate \citep{2007PhPl...14h2107H}. Null points are usually created and destroyed in pairs at a single point, and the absence of sub-grid structure in our model (see Section~\ref{sssec:stability}) means that they cannot be studied until they are far enough apart to be found in separate grid cells. However, this is not a significant issue since at the instant of creation, a new pair of nulls will separate at infinite speed \citep[][private communication]{priv:hornig09}, so a newly-created null pair will usually make its first appearance already separated by one or more grid cells, even if extremely short timesteps are used.

Once detected, magnetic null points can be classified into different types depending on the properties of their associated magnetic fieldlines \citep{1975RISRJ..29..133F,1988JGR....93.8583G,1996PhPl....3..759P}. Every generic 3D magnetic null point is the termination point for two singular spine fieldlines, set at a nonzero angle to a surface (planar close to the null) also consisting of fieldlines terminating at the null, called the separatrix surface (or, close to the null, the fan surface). To preserve the divergence-free condition, if the spine fieldlines point towards the null then the separatrix surface fieldlines must point away from it, and vice versa. The first type is called a \emph{positive} magnetic null point, the second type \emph{negative}. The fieldlines in the separatrix surface may point radially away from the null, which is then called \emph{proper}. A small amount of current parallel to the spine produces an \emph{improper} null in which some fan fieldlines experience a slight bend towards the major fan axis. If there is a current greater than a certain threshold flowing parallel to the spine, these fieldlines may take on a \emph{spiral} shape. A component of current perpendicular to the spine results in a different type of improper null where the spine and fan locally collapse towards one another, so that they are no longer perpendicular. In general, it is likely that there will be some current both parallel and perpendicular to the spine, resulting in both of these effects coming into play.

Positive/negative nulls may be distinguished by linearising the expression for the magnetic field vector close to the null, and considering the eigenvalues of the Jacobian matrix. Two will be of the same sign and one of the opposite sign; the two eigenvalues with the same sign give the sign of the null point. Spiral/improper nulls may also be distinguished by considering the eigenvalues and eigenvectors of the same Jacobian matrix. If they have an imaginary part, the null is spiral; if they are purely real, the null is improper.

\section{Results}
\label{sec:results}

\subsection{Magnetic null points}

\subsubsection{Number of nulls}

We restricted our search for magnetic null points to the regions of the model where the magnetic field vector has an appreciable nonzero value. The null-finding algorithm only works, and indeed it only really makes sense to talk about magnetic null points at all, in regions where the magnitude of the magnetic field vector is non-zero almost everywhere. In our model, there is initially a magnetic field throughout the whole atmosphere, and in the twisted flux tube in the convection zone. The weak-field region that was thus excluded consists of the portion of the model convection zone where the magnetic field strength is so weak as to be negligible. Including this region in the analysis would only have resulted in the inclusion of an enormous number of spurious null points, with no physical reality, but resulting purely from fluctuations of the order of the floating point accuracy of the code. We chose our cutoff value of magnetic field strength to minimise the number of these spurious nulls while keeping a very high sensitivity.

%We restricted our search for magnetic null points to the region of the model atmosphere where the horizontal background magnetic field exists. Below this level, there is no magnetic field initially except in the rising flux tube, so the value of the magnetic field vector is essentially zero everywhere. The null-finding algorithm only works, and indeed it only really makes sense to talk about magnetic null points at all, in regions where the magnitude of the magnetic field vector is non-zero almost everywhere. From now on, when we refer simply to ``the photosphere'', we are referring to the very top level of the photosphere, above which the horizontal background magnetic field is significant. For these reasons, we did not search for nulls below the photosphere, but we believe that, in any case, nulls can only exist (or at least are only significant) where fields with different magnetic connectivities interact, which in our model means above the photosphere.

The full flux emergence model has 220 time frames, covering 62 minutes in total. The rising flux tube begins to interact with the coronal magnetic field after about 20 minutes, and the first magnetic null points appear at about this time. Once the first nulls have appeared, more follow, and there are null points present in every frame from then on.

\begin{figure}
  \centering
  \parbox[t][0.35\textwidth][t]{\textwidth}{
    \raisebox{0.3\textwidth}{(a)}
    \includegraphics[width=0.45\textwidth]{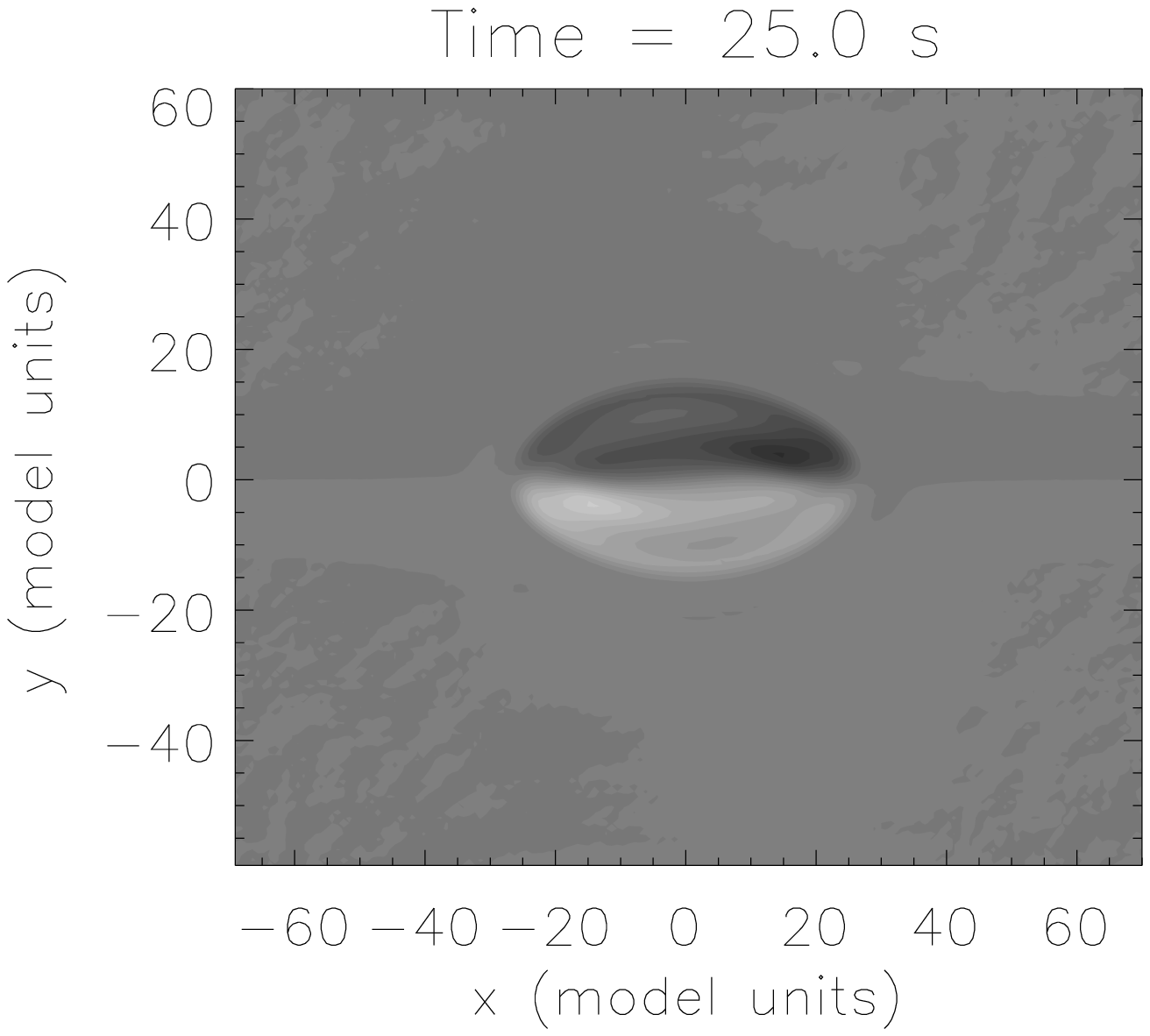}
    \raisebox{0.3\textwidth}{(b)}
    \includegraphics[width=0.45\textwidth]{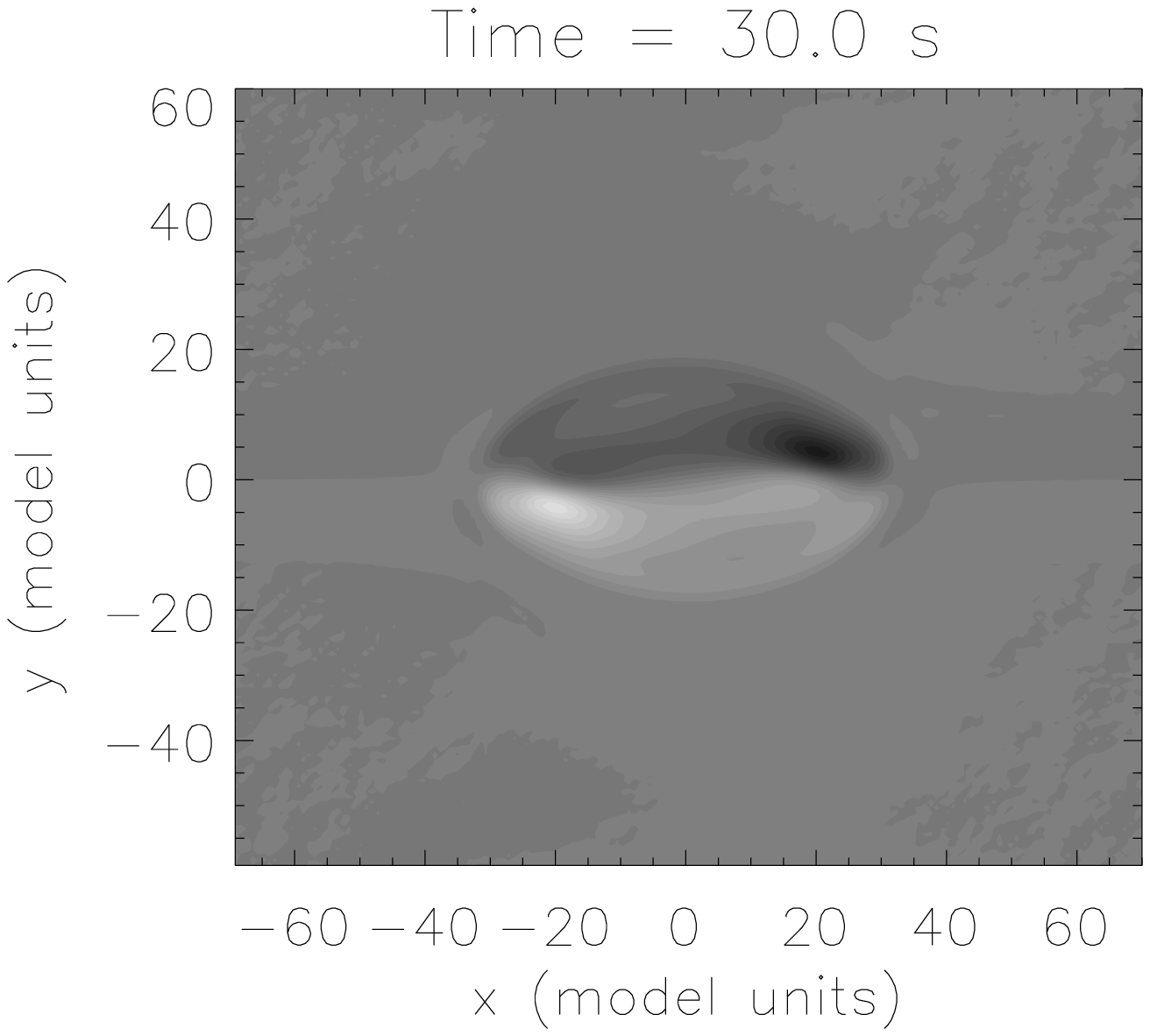}
  }
  \parbox[t][0.35\textwidth][t]{\textwidth}{
    \raisebox{0.3\textwidth}{(c)}
    \includegraphics[width=0.45\textwidth]{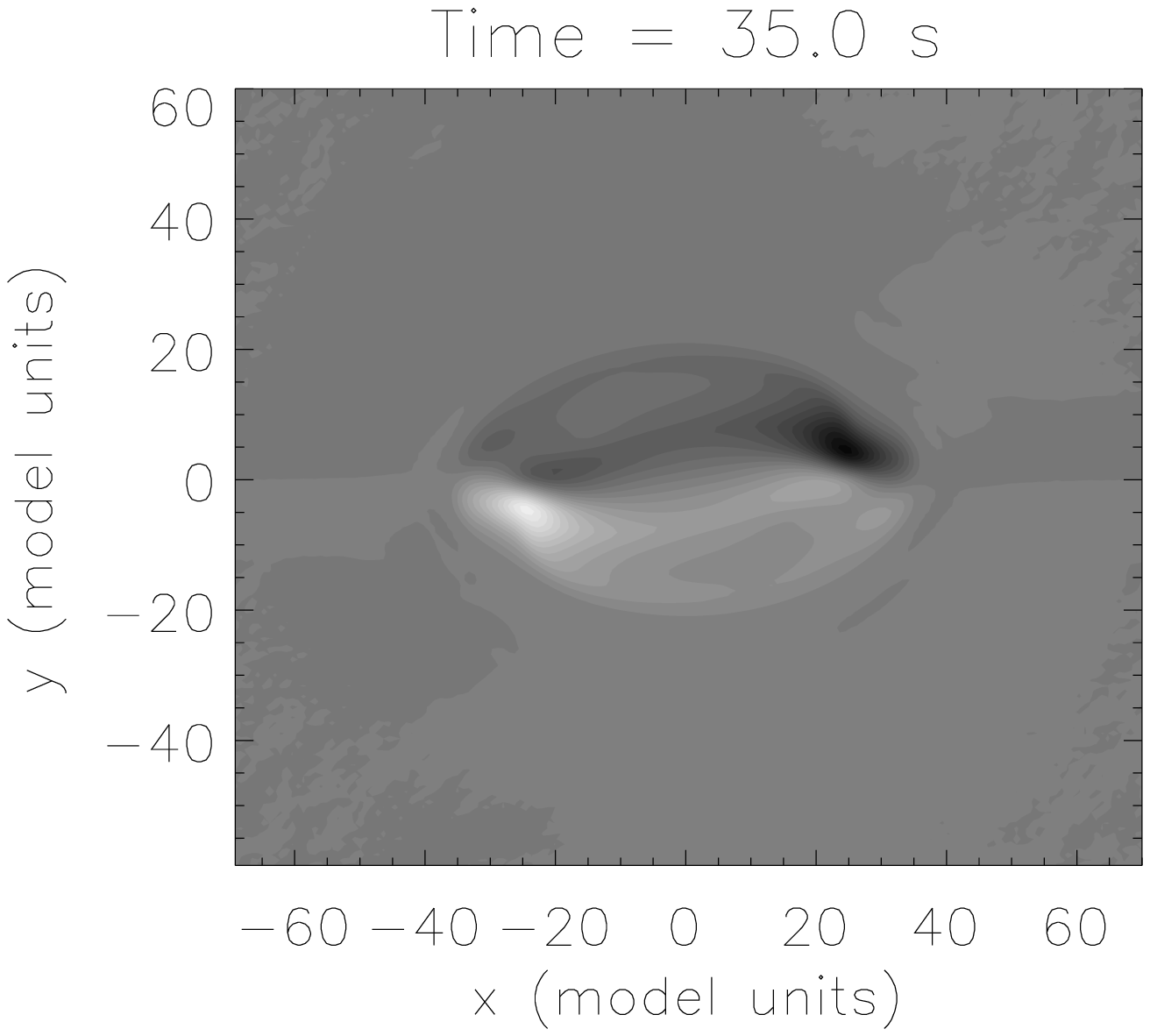}
    \raisebox{0.3\textwidth}{(d)}
    \includegraphics[width=0.45\textwidth]{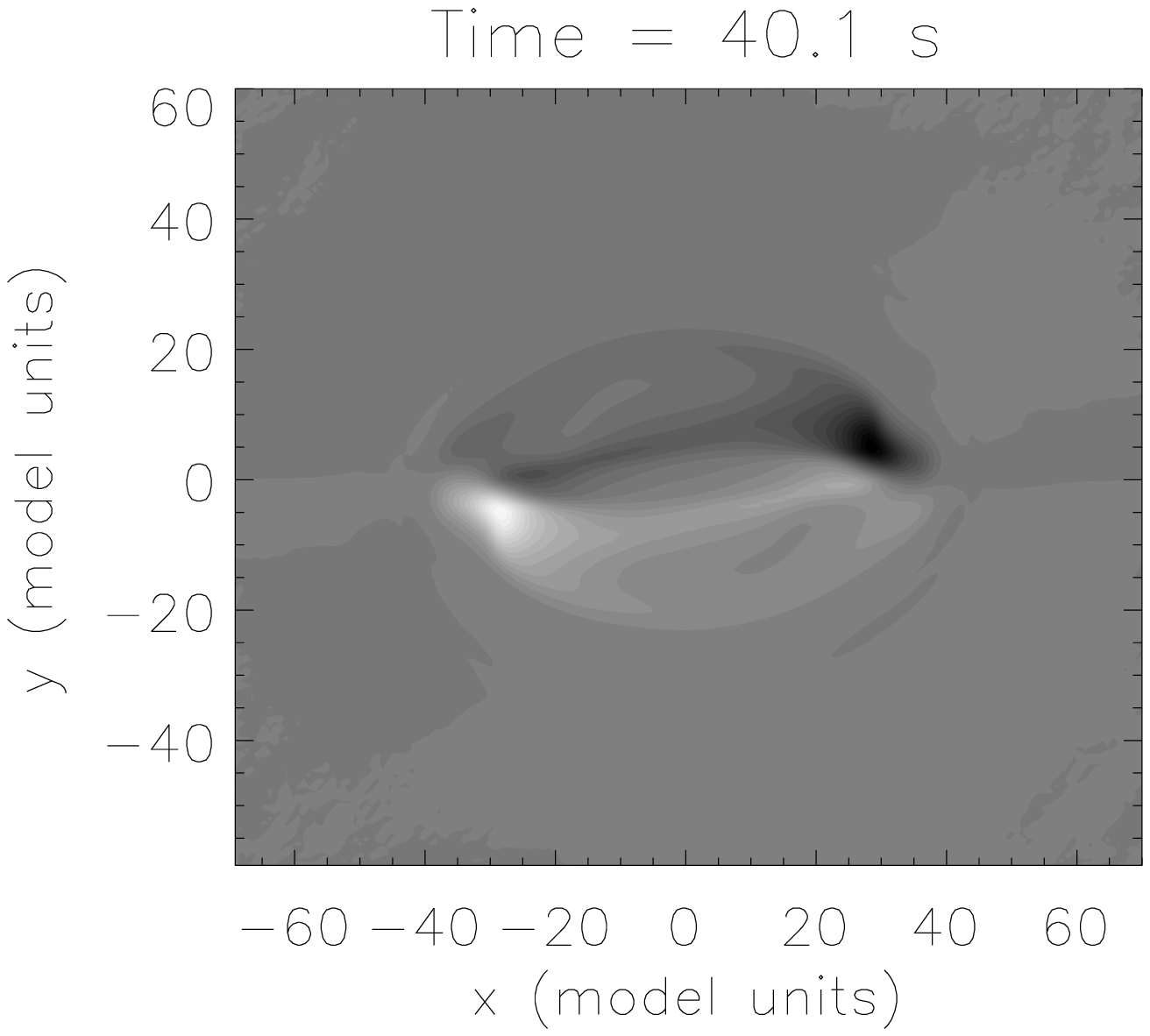}
  }
  \parbox[t][0.35\textwidth][t]{\textwidth}{
    \raisebox{0.3\textwidth}{(e)}
    \includegraphics[width=0.45\textwidth]{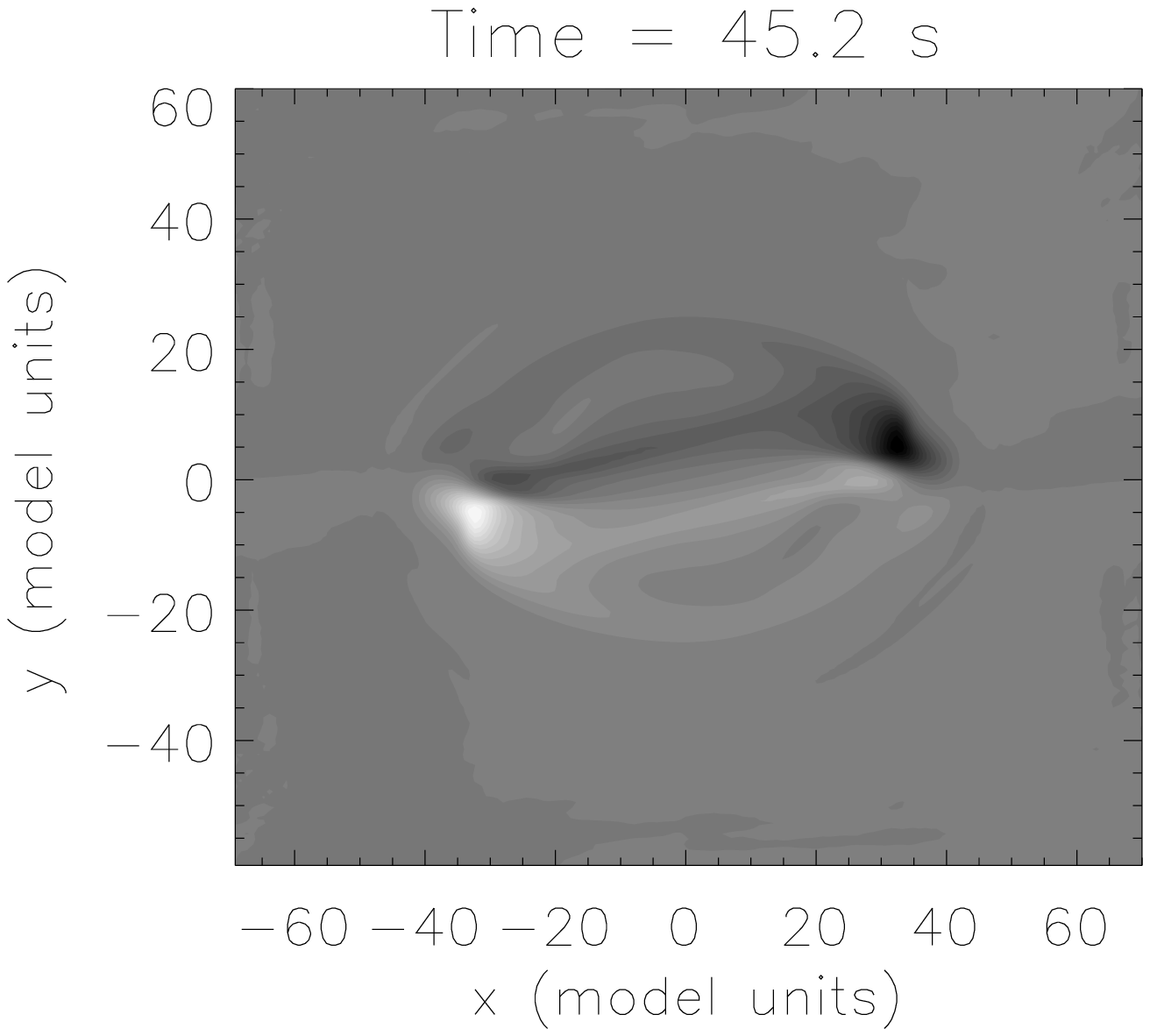}
    \raisebox{0.3\textwidth}{(f)}
    \includegraphics[width=0.45\textwidth]{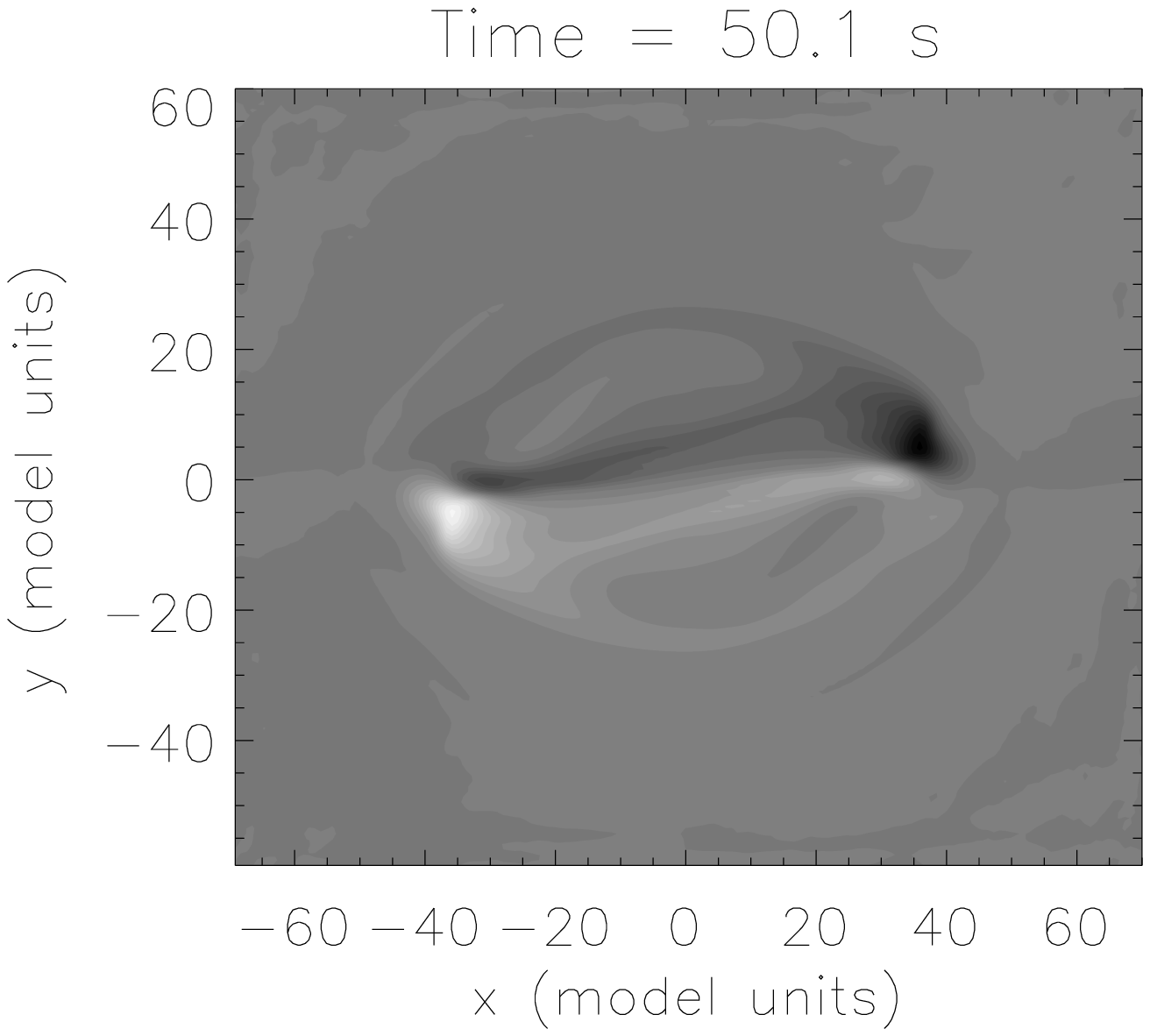}
  }
  \parbox[t][0.35\textwidth][t]{\textwidth}{
    \raisebox{0.3\textwidth}{(g)}
    \includegraphics[width=0.45\textwidth]{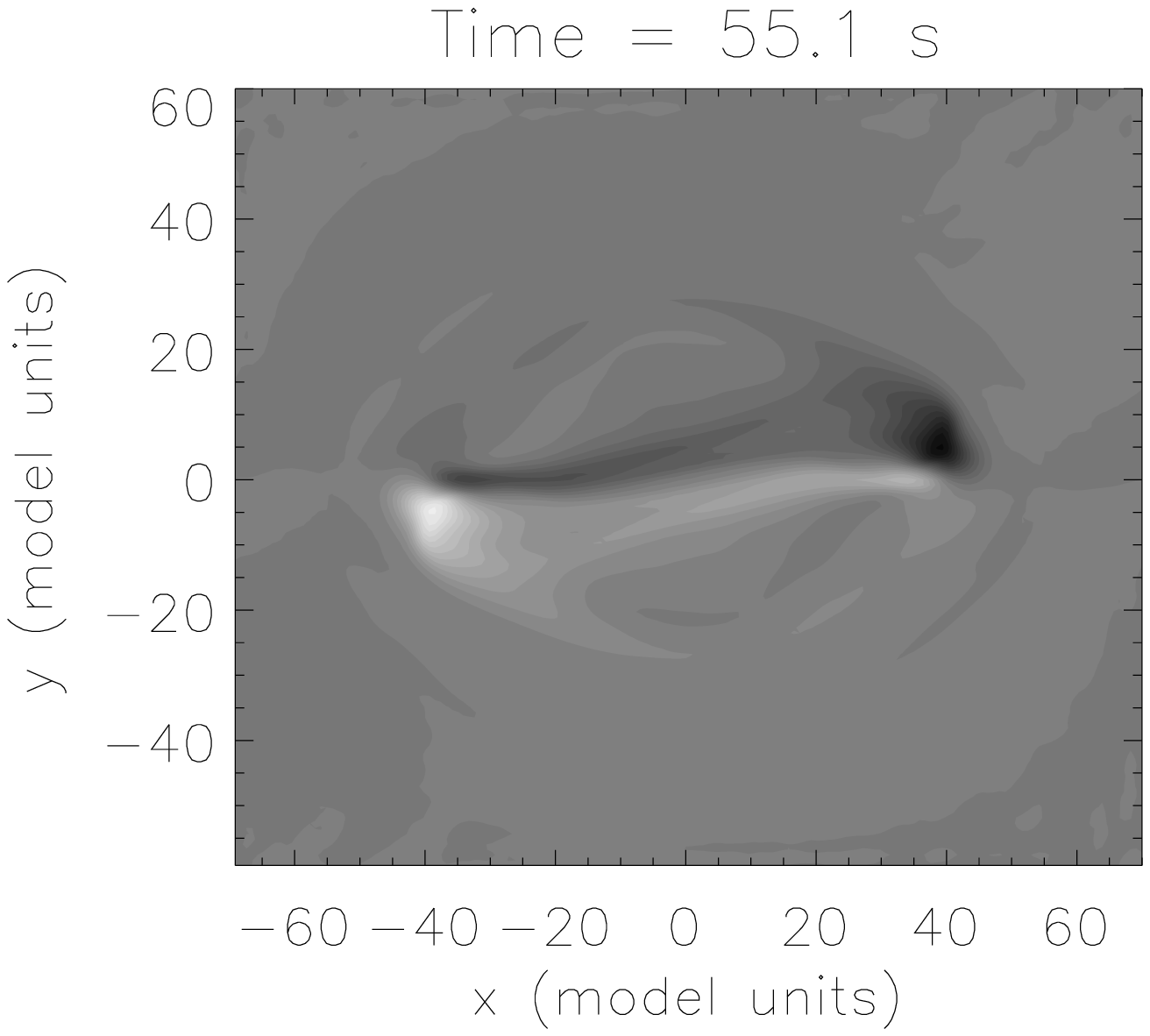}
    \raisebox{0.3\textwidth}{(h)}
    \includegraphics[width=0.45\textwidth]{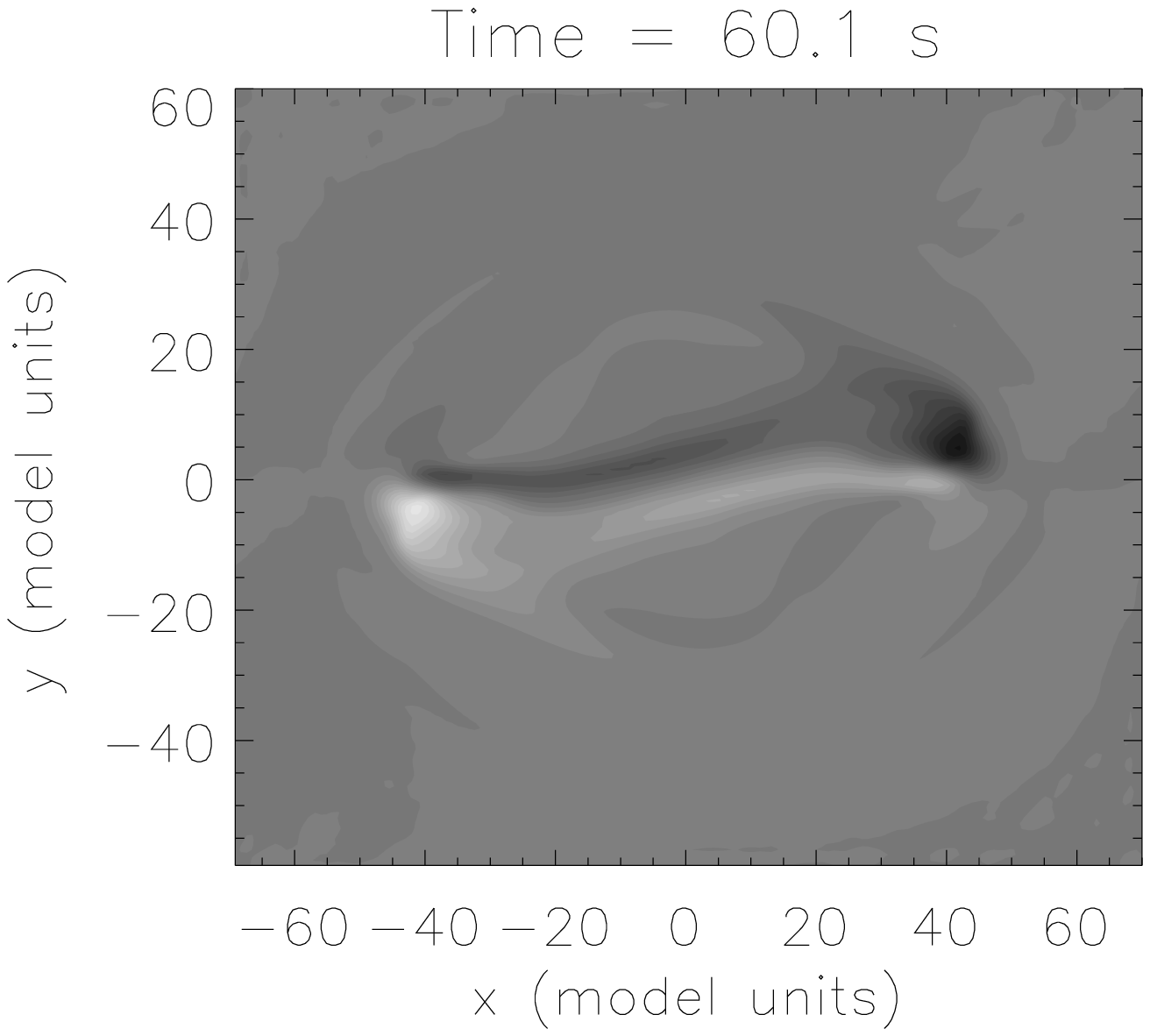}
  }
  \caption{Contour plots of the $z$-component of the magnetic field passing through the base of the photosphere at various times throughout the model run. The contour levels are normalised to the value of the strongest magnetic field in the final frame.}
  \label{fig:b_at_photo}
\end{figure}

Figure~\ref{fig:b_at_photo} shows snapshots of the component of the magnetic field vector normal to the base of the photosphere at different times throughout the model run. It is interesting that despite the obvious simplicity of the magnetic field passing through the photospheric ``boundary'', up to 26 magnetic null points are present in the model magnetic field at any one time. Throughout the model run, only two strong photospheric source regions of magnetic flux are present (one positive and one negative), and our previous experience of working with point-source potential-field magnetic topologies \citep[e.g.][]{2007SoPh..243..171M} would lead us to expect a maximum of one null point above the photosphere in such a configuration. However, magnetic fields that have undergone an MHD evolution, as opposed to a evolving in an equi-potential manner, can have a much higher degree of topological richness and complexity, as also found by \citet{2007RSPSA.463.1097H}. {Later, in Section~\ref{sec:conclusions}, we discuss possible reasons for the creation of so many nulls.}

\begin{figure}
  \centering
  \includegraphics[width=0.7\textwidth]{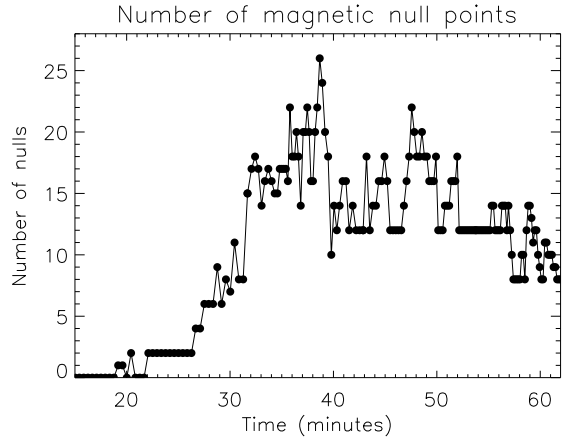}
  \caption{The number of magnetic null points present in the model magnetic field, as a function of time.}
  \label{fig:nnulls}
\end{figure}

The variation in the total number of nulls as a function of time is shown in Figure~\ref{fig:nnulls}. During the time period when nulls are present in the model, the mean number of nulls is $12.5$. The shape of the curve, with its high but short-lived peaks, combined with the fact that a large number of nulls are present for most of the period in question, suggests that there could be quite a large variation in the lifetimes of the nulls. This is investigated in Section~\ref{sssec:lifetimes}.

After the magnetic null points were identified, they were tracked from frame to frame \citep[][private communication]{priv:simpson08} using an algorithm which associates two nulls (n$_1$ and n$_2$) in consecutive time-frames (F$_1$ and F$_2$) if and only if:
\begin{enumerate}
\item n$_2$ is closer to n$_1$ than it is to any other null in F$_1$,
\item n$_1$ is closer to n$_2$ than it is to any other null in F$_2$,
\item the distance between n$_1$ and n$_2$ is less than some maximum tolerance (this is useful when the number of nulls in a frame is very small), and
\item n$_1$ and n$_2$ have the same sign.
\end{enumerate}
Counting every null in every frame separately, 2132 instances of nulls were detected by the null-finding code. The null-tracking allowed these nulls to be tracked and associated from one frame to the next, and in fact they boil down to a total of 305 nulls, each of which appears, moves around, and disappears in the course of the model run. The circumstances of their creation and destruction are discussed in the next section.

\subsubsection{Null point nature, creation, and destruction}
\label{sssec:nature}

\begin{figure}
  \centering
  \includegraphics[width=0.7\textwidth]{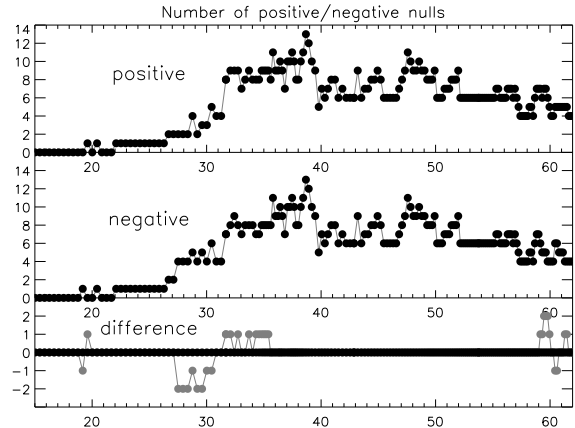}
  \caption{The number of positive and negative magnetic null points present in the model, as a function of time. Also plotted is the difference in the numbers of positive and negative nulls; the raw difference (grey) and the corrected difference (black; see further explanation in the main text).}
  \label{fig:nposneg}
\end{figure}

As described in Section~\ref{ssec:locnat}, magnetic null points can be classified by their sign and type, depending on the eigenvalues of the Jacobian matrix describing the linearised magnetic field near each null. Figure~\ref{fig:nposneg} shows how the numbers of positive and negative nulls vary in time in our flux emergence model, and the difference between the two.

The number of source regions and null points in a magnetic field must always obey the 3D Euler characteristic equation \citep{1999SoPh..186...99I}
\begin{equation}
S_{+} - n_{+} = S_{-} - n_{-}
\label{eq:euler}
\end{equation}
for positive and negative sources ($S_{\pm}$) of magnetic field and positive and negative magnetic null points ($n_{\pm}$). While the nulls are present in our model, the number of source regions remains fixed, and we begin with no magnetic null points whatsoever, so the equation says that the number of positive and negative nulls should remain balanced throughout the model run. As Figure~\ref{fig:nposneg} shows, this is the case nearly all of the time, except for a few isolated occasions when the raw numbers of positive and negative nulls are not balanced, the reasons for which are explained below.

Magnetic null points are created and destroyed by topological bifurcations of the magnetic field. Several types of topological bifurcation can create and destroy nulls, but they must all respect the 3D Euler equation and preserve the balance between positive and negative null points. The simplest bifurcation that can do this is the local separator bifurcation \citep{1999RSPSA.455.3931B}, which creates or destroys a pair of oppositely-signed magnetic null points. We have individually studied the birth and death of each null point in the model. For the vast majority of the nulls, their circumstances of creation and destruction are consistent with a known topological bifurcation of the magnetic field, and the balance of signs from the Euler equation is preserved. However, there is a very small number of exceptions to this rule, comprising about $1.8$\% of the total number of nulls detected.

The imbalances in the number of positive and negative nulls occur in each case when an isolated null appears to be created without an oppositely-signed partner. This always takes place in an extremely weak-field region, in fact at the boundary of the weak-field cutoff region described previously. An isolated null can occasionally appear in this vicinity for one of two reasons. The first possibility is that it is a real null, but its partner has been created inside the cutoff region and therefore cannot be detected. Such isolated nulls may themselves be detected and then not detected in successive frames, due to their proximity to the weak-field cutoff region. Our imposition of a strict cutoff level means that as the model evolves in time, different grid cells will be ruled in and out of our detection region, and if these grid cells contain nulls, the nulls could seem to appear and disappear. The other possibility for the detection of an isolated null is that the null is caused by floating point errors, as is sometimes possible so close to the weak-field cutoff region. We chose our cutoff level at $1.3 \times 10^{-4} \, \mathrm{G}$ ($1 \times 10^{-7}$ in model units), because this level is low enough to retain sufficient sensitivity to detect real nulls in quite weak magnetic fields, but high enough to minimise detections of insignificant nulls caused by floating point errors in extremely weak-field regions.
% As explained previously, it is meaningless to talk about finding null points in regions where the magnetic field is so incredibly weak as to be essentially zero everywhere. In such a region, the null-finding algorithm would still work perfectly, but the null points that it would return are in some sense not real or meaningful, and are certainly not interesting for our analysis, so we exclude them by limiting our detection region to grid cells whose magnetic field strength is over the cutoff level almost everywhere.

We have examined in detail each frame in which there is an imbalance in the number of positive and negative nulls. The first period of imbalance, before 20 minutes, is caused by the detection of false nulls due to floating point errors, as is the final period at around 60 minutes. The imbalance between 27 and 36 minutes, are caused by isolated nulls very close to the cutoff region, whose partners must be inside. We can distinguish between the two possibilities because a real null will show stability (see Section~\ref{sssec:stability}) as it is tracked from one frame to the next, while a false null caused by floating point errors in a very weak magnetic field only appears in one frame, and is often far away from the other real nulls that exist at that time.

These results are strong evidence for the reality of the magnetic null points that we have detected in the model magnetic field. They never appear or disappear without a logical cause; in every single case, their creation or destruction is caused either by a known topological bifurcation or can be explained by their proximity to the weak-field cutoff region. Also, the number of positive and negative nulls present always obeys the conservation law derived from the 3D Euler equation, once spurious and isolated nulls have been accounted for.

An interesting point to note is that the sign of a null is also conserved in the sense that a null point sitting at a particular location is never seen to switch sign from frame to frame. Clearly according to the tracking algorithm, it would not be identified as the same null if it did show this behaviour, but watching a movie of the nulls (see supplementary online material) makes it plain that such behaviour does not occur anyway.

The nulls can also be classified according to their type (improper or spiral), and since this is purely dependent on the electric current density at the null, the type of a null can (and does) freely change in time. Some of the nulls in our experiment change from improper to spiral and vice versa several times in their tracked lifetime. Spiral nulls make up about 30\% of the total population of nulls in our experiment.

\begin{figure}
  \centering
  \parbox[t][0.35\textwidth][t]{\textwidth}{
    \raisebox{0.4\textwidth}{(a)}
    \includegraphics[width=0.45\textwidth,clip=true]{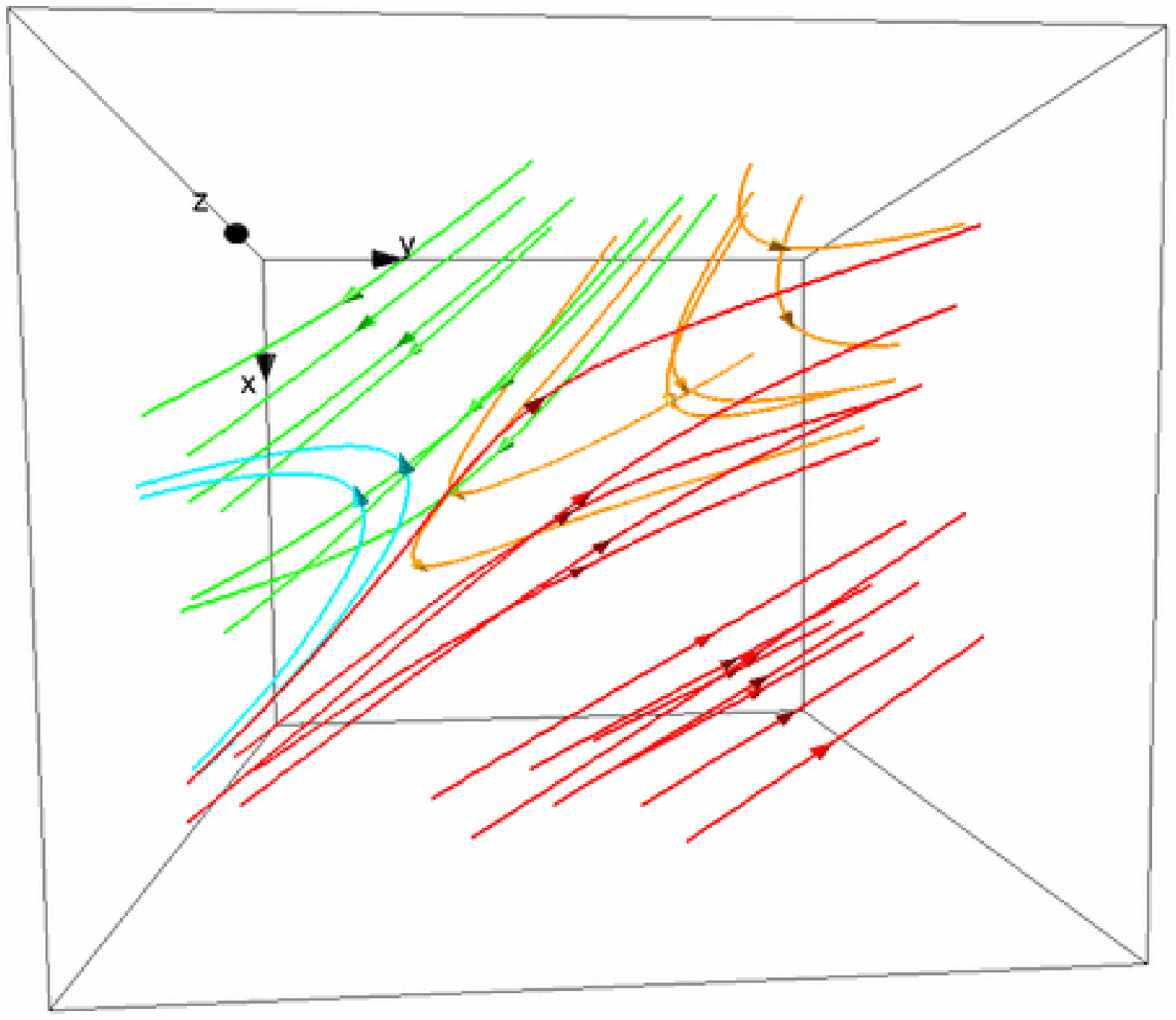}
    \raisebox{0.4\textwidth}{(b)}
    \includegraphics[width=0.45\textwidth,clip=true]{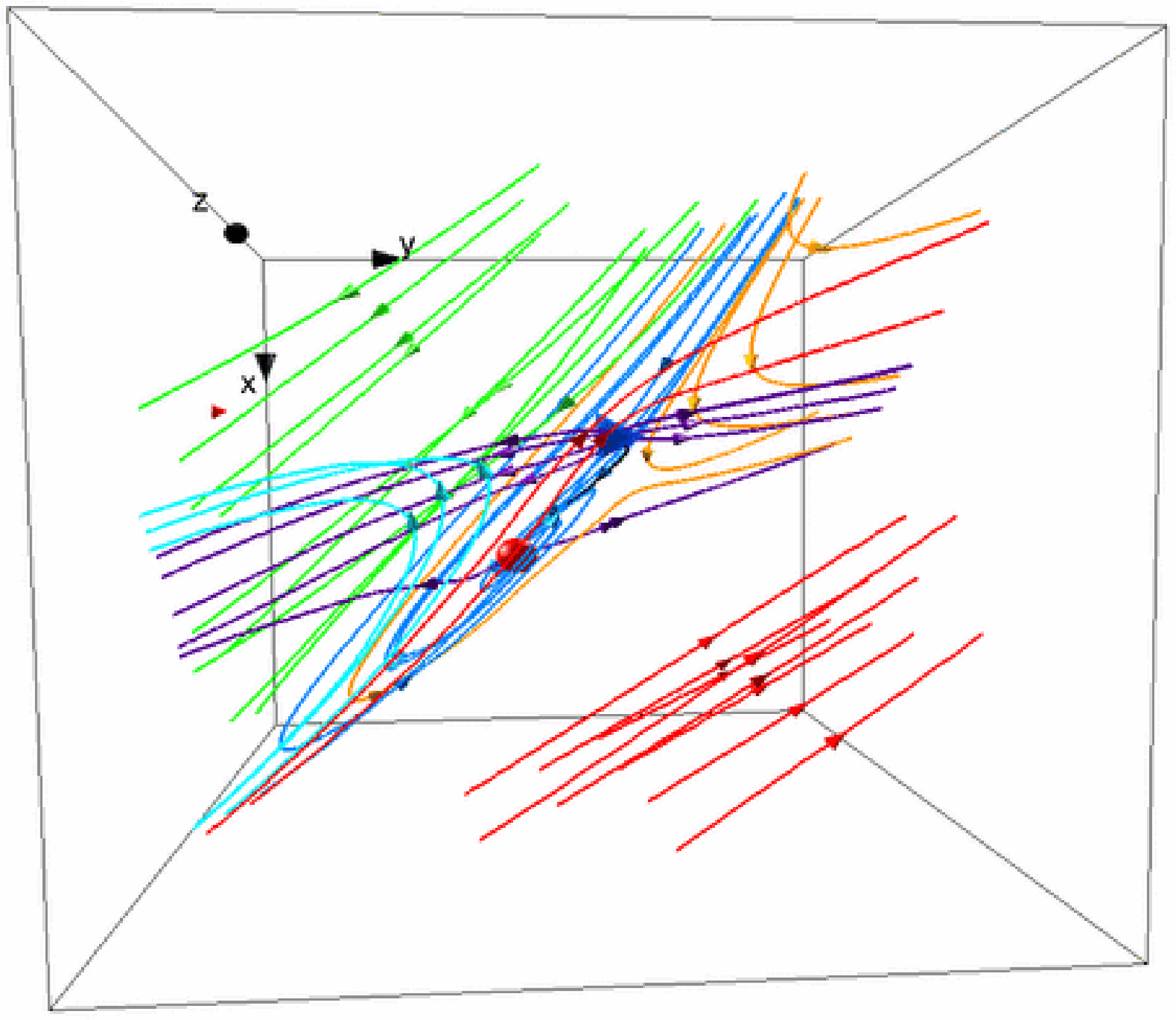}
  }
  \parbox[t][0.35\textwidth][t]{\textwidth}{
    \raisebox{0.4\textwidth}{(c)}
    \includegraphics[width=0.45\textwidth,clip=true]{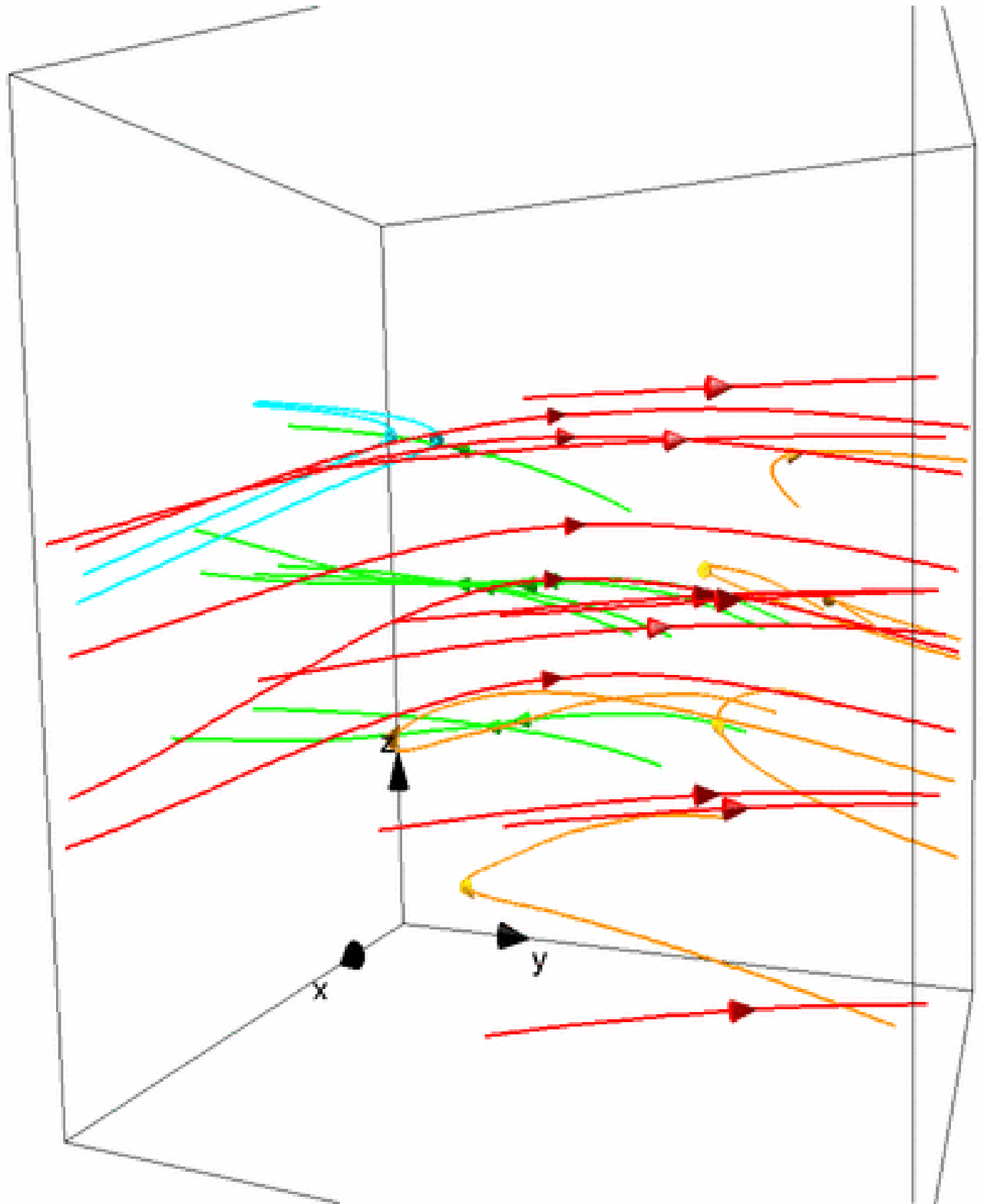}
    \raisebox{0.4\textwidth}{(d)}
    \includegraphics[width=0.45\textwidth,clip=true]{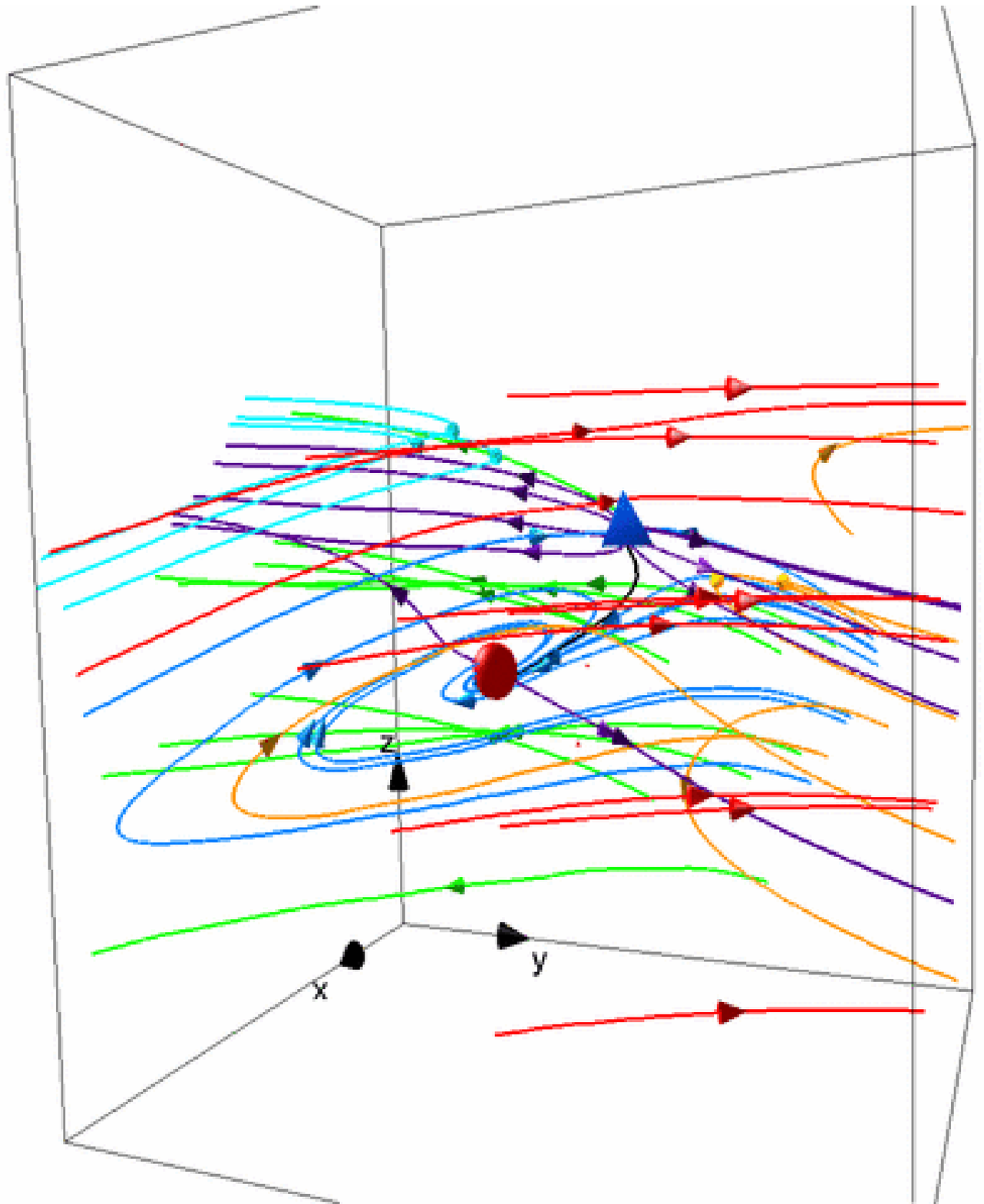}
  }
%   \includegraphics[width=0.47\textwidth]{before_colours_plan.eps}
%   \hfill
%   \includegraphics[width=0.47\textwidth]{after_colours_plan.eps}
%   \includegraphics[width=0.47\textwidth]{before_colours_side.eps}
%   \hfill
%   \includegraphics[width=0.47\textwidth]{after_colours_side.eps}
  \vspace{0.5\baselineskip}
  \caption{3D views of the magnetic fieldlines in a small cube surrounding the location where a new pair of nulls appears in the atmosphere. The cube has an edge length of 1700km, which corresponds to about 12 grid cells in the horizontal directions and 35 grid cells in the vertical direction. Two sets of snapshots are shown, from consecutive time frames in the model; (a) and (c) are just before the nulls were created, at 43 minutes and 31 seconds, and (b) and (d) are just afterwards, at 43 minutes and 49 seconds. (a) and (b) show the fieldline structures from above, and (c) and (d) show side views of the same fieldlines. The fieldlines are colour-coded by their direction, which is also indicated by arrows. Green fieldlines run from high y-values to low y-values; red fieldlines run the opposite way, from low y to high y. Yellow fieldlines curve back on themselves in y, running from low x to high x; cyan fieldlines complete the set, also curving back on themselves to run from high x to low x. The yellow fieldlines are generally lower in z than the cyan ones. The null pair in (b) and (d) consists of a positive spiral null (shown as a red sphere) and a negative improper null (a blue tetrahedron). A separator fieldline, shown in black, joins the newly-formed nulls. Each null also has its associated spine and separatrix surface fieldlines plotted. Since the separatrix surface of each null is constrained to run along the spine of its partner, the spines of the improper null and the separatrix fieldlines of the spiral null are both shown in blue, and vice versa in purple for the other set of spines and separatrix fieldlines.}
  \label{fig:f123124}
\end{figure}

Figure~\ref{fig:f123124} shows before (a \& c) and after (b \& d) snapshots of the creation of a new pair of nulls in the atmosphere. The size of the box where the fieldlines are plotted is 1700km along each edge, corresponding to about 12 grid cells horizontally and 35 grid cells vertically. We have studied the fieldline structure around many pairs of newly-formed nulls, and this is a typical example. Fieldlines pointing in four different directions, colour-coded and with arrows to show direction, are present in the vicinity both before and after the null pair is formed. The views from above (a \& b) are reminiscent of a classical 2D reconnection scenario, but the side views (c \& d) make it clear that we are dealing with a fully 3D magnetic structure. The two newly-formed nulls are joined by a separator.

In fact, an isolated newly-formed pair of nulls must be linked by at least one separator \citep{1990ApJ...350..672L}. In general, this means that the separatrix surface of each null extends out to touch the other null, and is bounded by the spine of this null, so that the two surfaces intersect along a common separator. The MHD formation process of null points has not been explored in depth, although recently \citet{otto2009} has run a experiment in which an isolated null pair is formed. It is extremely unlikely that any MHD numerical experiment will ever be able to capture the exact instant of creation of a pair of nulls, as the nulls move apart infinitely fast (for an infinitesimal period of time) immediately upon their formation \citep[][private communication]{priv:hornig09}, thus gaining a finite separation infinitely quickly. Indeed, the experiment of \citet{otto2009}, which focusses simply upon the creation of the null pair, naturally has higher spatial and temporal resolution in the vicinity of the null pair than we have, and it clearly shows this behaviour. In our experiment, the first detection of a new pair of nulls can show two nulls with any combination of types, i.e.\ two improper nulls, an improper and a spiral, and two spirals. But as explained above, it is impossible to tell whether all the pairs are created as two improper nulls, of which some then quickly gain a large enough current density to become spiral, or whether new spiral nulls are indeed being directly created in the topological bifurcations.

\begin{figure}
  \centering
  \includegraphics[width=0.7\textwidth]{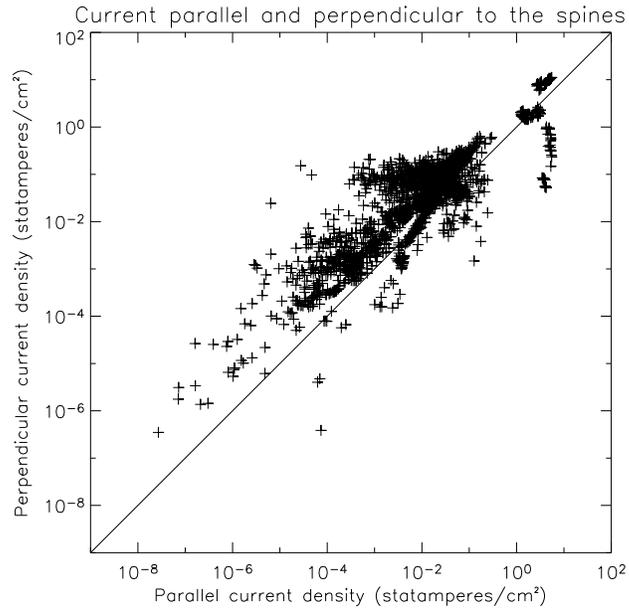}
  \caption{Scatter plot of the absolute value of current density at each of the magnetic null points, parallel and perpendicular to their spine fieldlines. We have used logarithmic axes in x and y in order to capture the full range of the current densities, which cover more than eight orders of magnitude.}
  \label{fig:nimpspir}
\end{figure}

However, in some ways, classifying a null as spiral or improper is just an imprecise way of looking at the current density close to it. Figure~\ref{fig:nimpspir} is a scatter plot showing the current density at every null in the model, both parallel and perpendicular to their spine fieldlines. There is a huge range of current densities at the different nulls. A strong parallel current density causes the fieldlines in the separatrix surface to spiral, and a strong perpendicular current density causes the angle between the spine and the separatrix surface close to the null to decrease. A small proportion of the nulls are dominated by either parallel or perpendicular current densities, which implies that they are close to being pure spiral or improper nulls respectively. However, most of the nulls have fairly similar values for their parallel and perpendicular current densities, which means that both effects come into play in determining their type and local fieldline structure.

\subsubsection{Lifetimes of nulls}
\label{sssec:lifetimes}

\begin{figure}
  \centering
  \includegraphics[width=0.7\textwidth]{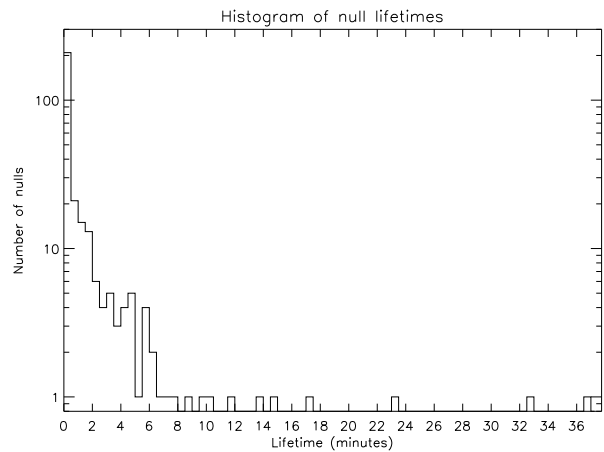}
  \caption{Histogram of null lifetimes. The mean timestep between frames is 17 seconds.}
  \label{fig:lthist}
\end{figure}

Figure~\ref{fig:lthist} is a histogram of the lifetimes of the null points. When interpreting this histogram, it is important to consider the length of the time step between frames. The actual time step used in the code is variable, in order to maintain an acceptable balance between efficiency and accuracy. Data is read out from the code at intervals containing many of these short time steps. Each readout of data is referred to as a ``frame''. The mean time step between frames is 17 seconds, and the maximum is 25 seconds. 14 of our 305 nulls flit in and out of existence, only appearing in one frame, but the evidence from Section~\ref{sssec:nature} strongly supports the reality of even these nulls. Most of the nulls persist for several frames or longer, and the longest-lived one survives for almost half of the entire time period during which nulls are present in the model.

\subsubsection{Stability of nulls}
\label{sssec:stability}

Another way of confirming the reality of the nulls is to look at how the directions of the eigenvectors corresponding to their spines changes in time. An orientation changing randomly from frame to frame would suggest that it is not in fact one null being tracked, but a sequence of different nulls. On the other hand, a steady or slowly-changing spine direction suggests that the null possesses a certain stability and retains its identity.

\begin{figure}
  \centering
  \includegraphics[width=0.7\textwidth]{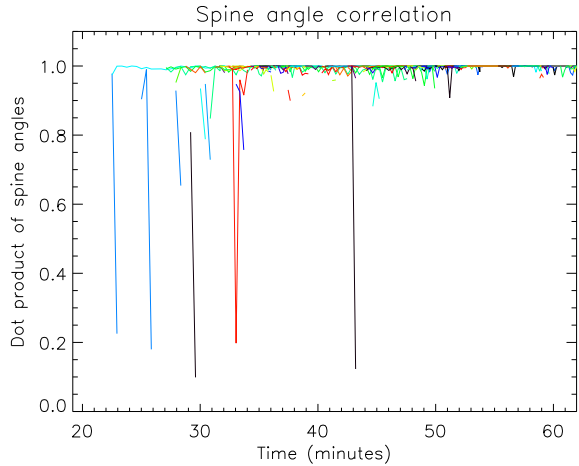}
  \caption{A measure of the stability of the nulls, based on the time variation of their spine angles. A value close to 1 means that the null's spine direction varies only very slowly in time, and hence the null is stable.}
  \label{fig:spineangle}
\end{figure}

Figure~\ref{fig:spineangle} is a plot showing how each null point's spine direction changes in time, for all of the nulls that last two or more frames. Each null is represented by a different coloured line, and the quantity plotted is the dot product of the normalised spine eigenvectors for the previous and current frames, at each time when the null exists. It is clear that most of the nulls show a remarkable degree of stability in their spine directions, shown by the fact that the eigenvectors from one frame to the next are very close to being parallel. 94\% of all the spine pairs have a correlation factor larger than $0.9$.

\begin{figure}
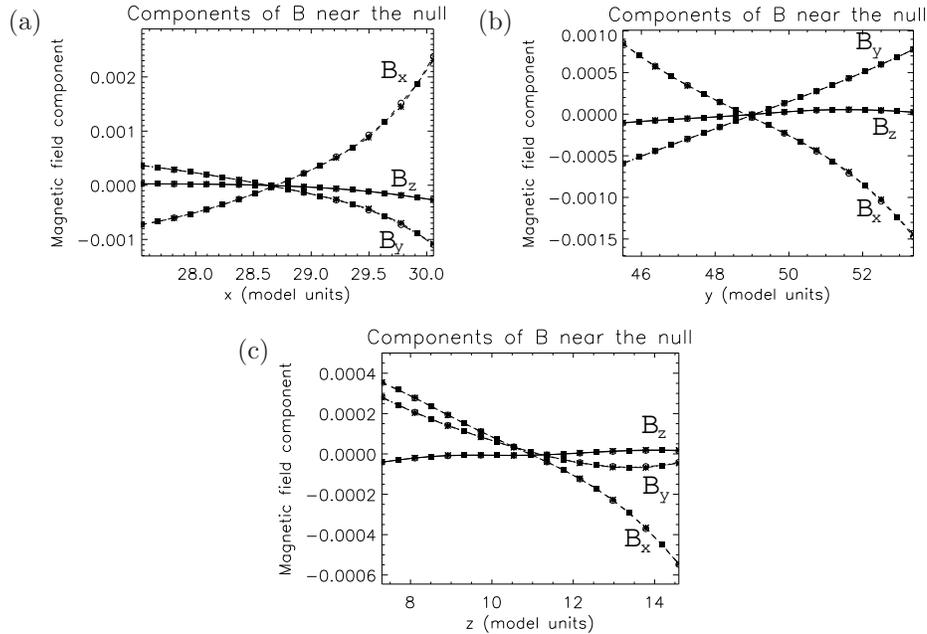

  \centering
  \parbox[t][0.35\textwidth][t]{\textwidth}{
    \raisebox{0.3\textwidth}{(a)}
    \includegraphics[width=0.43\textwidth,clip=true]{bxvar_00114_11.epsi}
    \hfill
    \raisebox{0.3\textwidth}{(b)}
    \includegraphics[width=0.44\textwidth,clip=true]{byvar_00114_11.epsi}
  }
  \vspace{0.25\baselineskip}
  \parbox[t][0.35\textwidth][t]{0.5\textwidth}{
    \raisebox{0.3\textwidth}{(c)}
    \includegraphics[width=0.45\textwidth,clip=true]{bzvar_00114_11.epsi}
  }
  \caption{Variation of the $\mathbf{B}_{x}$, $\mathbf{B}_{y}$, and $\mathbf{B}_{z}$ components of the magnetic field as functions of x, y, and z for a randomly selected null point from the model. Square plotting symbols indicate the original staggered data, circles indicate the simple linear interpolation from the staggered data, and asterisks indicate the high-order interpolated data that we used for the calculations throughout the paper.}
  \label{fig:bxyzvar}
\end{figure}

We can also study the spatial stability of the nulls by checking that the high-order nature of the code does not produce ringing or overshoots, and that our interpolation from the original staggered grid onto a centred grid has not introduced any other spurious oscillations in the magnetic field. Figure~\ref{fig:bxyzvar} shows plots of the variation with $x$, $y$, and $z$ of $\mathbf{B}_{x}$, $\mathbf{B}_{y}$, and $\mathbf{B}_{z}$ from the original staggered data, from linearly interpolating onto a centred grid, and from high-order interpolation onto a centred grid. We use the latter throughout this paper. Although the plots show three sets of points for $\mathbf{B}_{x}$, $\mathbf{B}_{y}$, and $\mathbf{B}_{z}$, the differences between them are so small that they are almost impossible to see. These plots are for a randomly selected null at a time of 40 minutes and 54 seconds into the model run. We have created plots like these for many of the nulls in the model, and they all show that both of the interpolated and the staggered curves lie practically on top of each other everywhere, and that all three curves always go through zero at the location of the null. Furthermore, if Gibbs overshoots were occurring in the model magnetic field, they should be found wherever $\mathbf{B}$ has a strong gradient (i.e.\ where $\mathbf{j}$ is largest), but clearly this is not the case as our nulls occur in just two clusters rather than, for instance, all along the edge of the flux tube below the photosphere (see Section~\ref{ssec:distmove}). Furthermore they only arise once the flux tube starts to emerge and interact with the atmospheric magnetic field.

\subsubsection{Distribution and movement of nulls}
\label{ssec:distmove}

The nulls are located in two loose clusters at either end of the emerging flux tube. They start to appear at the same time as when the magnetic reconnection between the emerging flux tube and the overlying field begins, i.e.\ when the rising flux tube has just touched the overlying field for the first time, and is starting to interact with it. The first real nulls appear at 22 minutes, and the reconnection begins at the same time. The nulls are located (and remain) close to the boundary between the four different magnetic connectivities.

\begin{figure}
  \centering
  \parbox[t][0.35\textwidth][t]{0.65\textwidth}{
    \raisebox{0.4\textwidth}{(a)}
    \includegraphics[width=0.6\textwidth,clip=true]{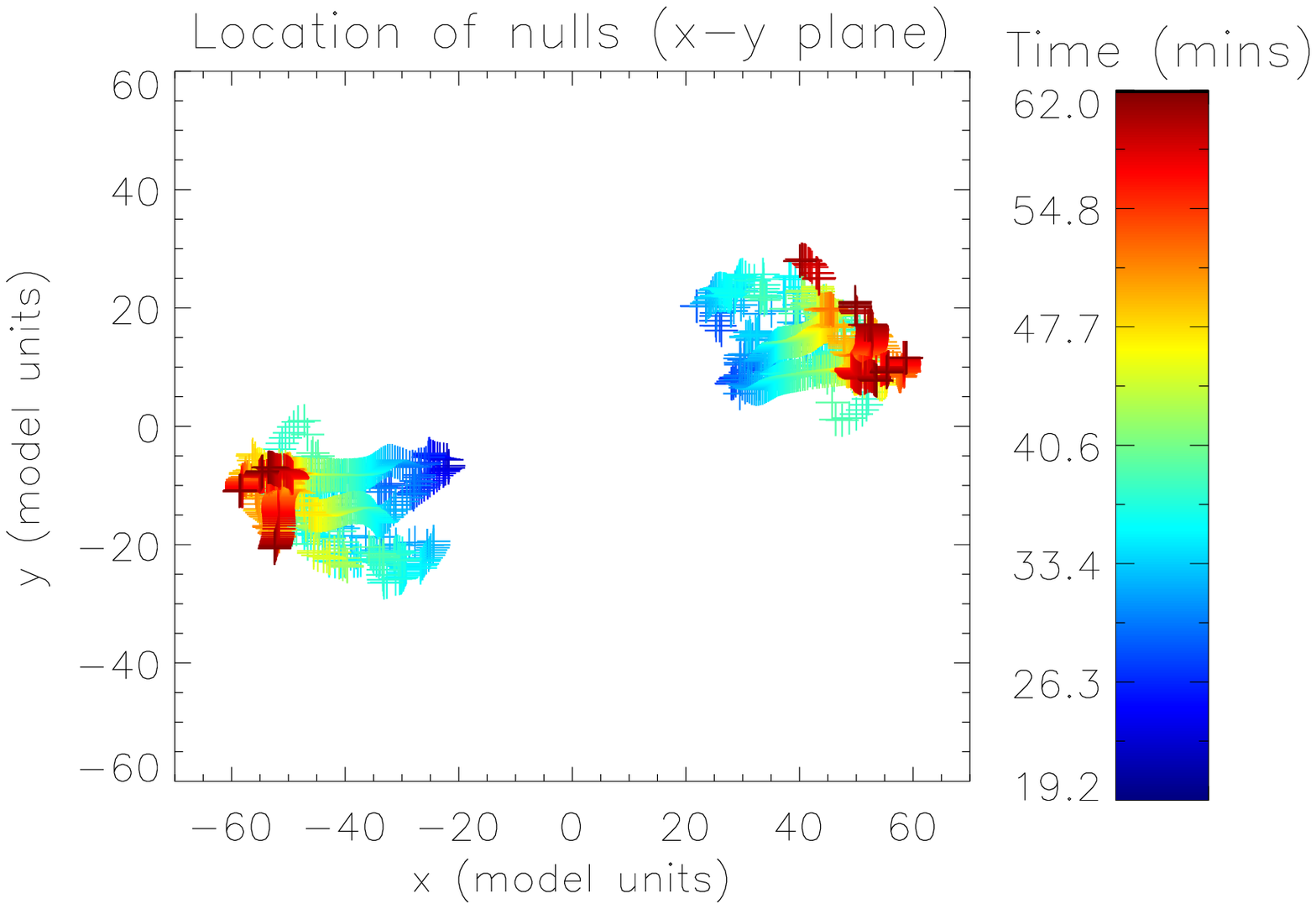}
  }
  \parbox[t][0.35\textwidth][t]{\textwidth}{
    \raisebox{0.275\textwidth}{(b)}
    \includegraphics[width=0.45\textwidth,clip=true]{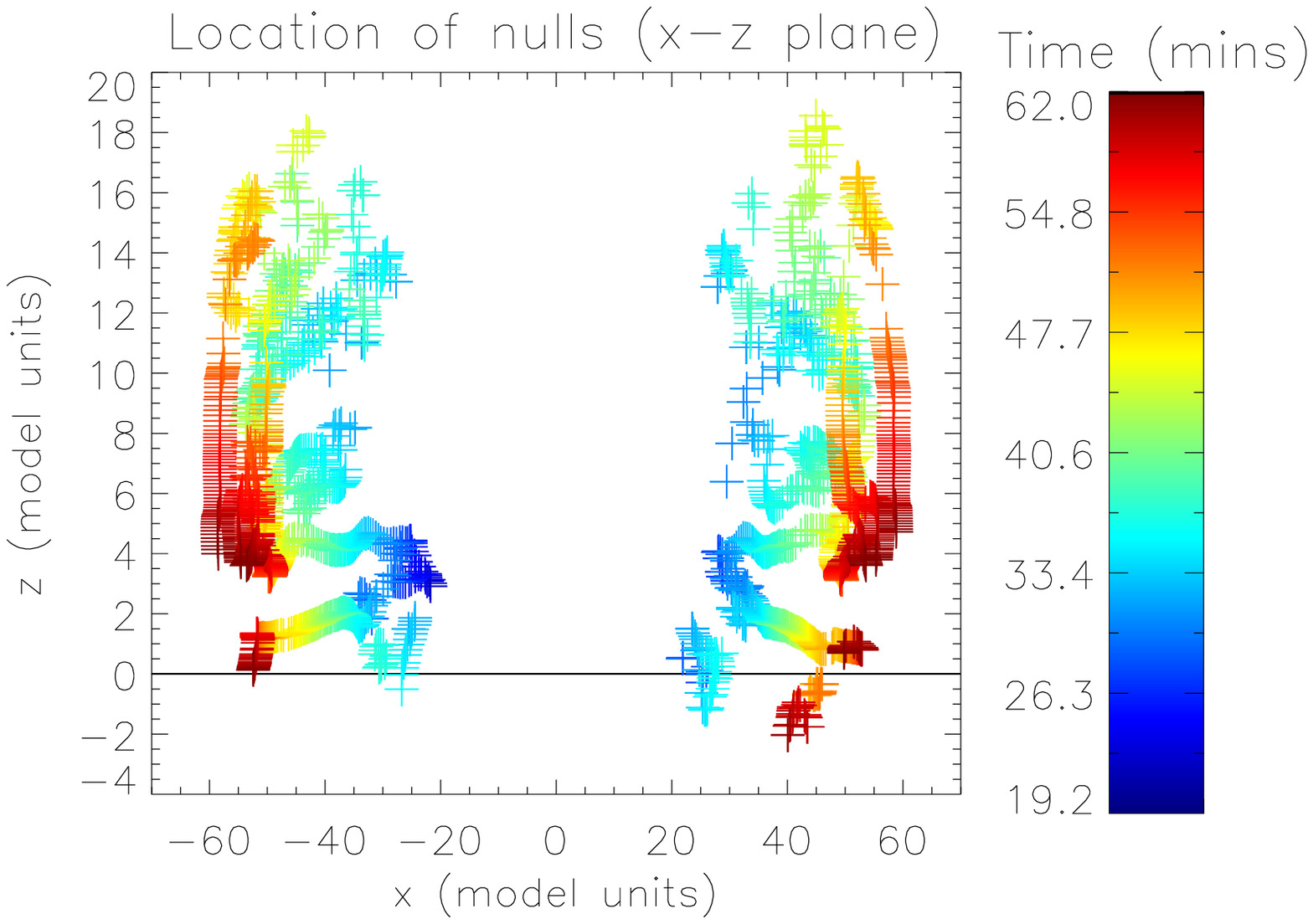}
    \raisebox{0.275\textwidth}{(c)}
    \includegraphics[width=0.45\textwidth,clip=true]{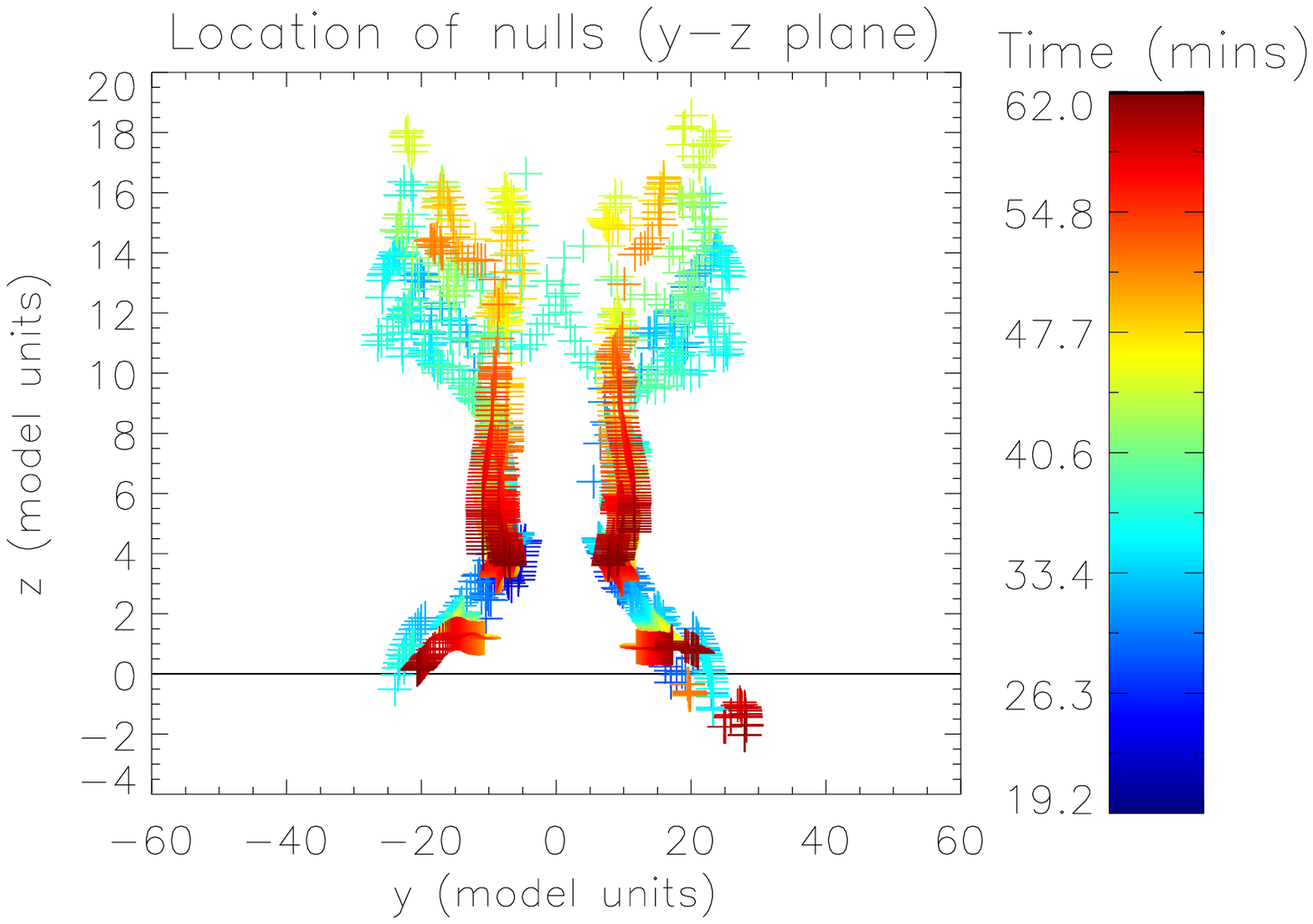}
  }
%   \includegraphics[width=0.6\textwidth]{new_xynulls.eps}
%   \includegraphics[width=0.48\textwidth]{new_xznulls.eps}
%   \hfill
%   \includegraphics[width=0.48\textwidth]{new_yznulls.eps}
  \caption{x-y, x-z, and y-z projections of the exact locations of the magnetic null points as they evolve in time. The colour of each null indicates at which time it existed. The black horizontal lines mark the position of the photosphere.}
  \label{fig:xyznulls}
\end{figure}

Figure~\ref{fig:xyznulls} shows, in the three different projections, how the exact locations of all the nulls vary in time, as found by using the final positioning step in the null-finding algorithm. The nulls first appear low in the photosphere, in two clusters at either side of the rising flux tube. The clusters expand upwards as more nulls appear at the edge of the flux tube. As the flux tube expands and reconnects with the overlying magnetic field, the nulls move out towards the side boundaries of the numerical box, although they always stay at least 5 model units away from the side boundaries and never move through them.  The nulls also move higher up through the photosphere and into the transition region. They peak at a height of $3.16\,$Mm, which means that they very nearly reach the top of the model transition region but never make it up as high as the model corona. As time goes on, they move back down towards the photospheric base, and the two clusters of nulls also continue to move apart as the flux tube continues to expand. Throughout this whole evolution there are always nulls in the low photosphere, and towards the end of the experiment this is the only region where any nulls remain.

The balance between positive and negative nulls in the two null clusters provides some important clues about the overall topology of the magnetic field, and in particular, where separators must exist. It is important to realise that the null pairs do not form somewhere in the middle above the emerging flux tube and then spread out to either side of the emerging flux region. In fact, the two clusters operate independently in terms of null point creation and destruction. Pairs of nulls are created and destroyed in each cluster without reference to the other cluster, which keeps the signs of the nulls in each cluster balanced. The only imbalances that are created, as described earlier, are due to isolated spurious nulls which are not created or destroyed in accordance with equation~\ref{eq:euler}. As we have already justified disregarding these nulls, they have not been included in Figure~\ref{fig:xyznulls}.

This might seem to imply that no separator(s) can join the two null clusters. However, this conclusion is wrong. The work of \citet{2007RSPSA.463.1097H} clearly showed that new separators can form between two pre-existing nulls that are far apart, via the global double-separator bifurcation. Indeed, it has been known for some time that this can also happen via, for example, the global spine-fan bifurcation or the global separator bifurcation \citep{1999RSPSA.455.3931B,2002SoPh..209..333B}. In our experiment, it is quite possible that at least one but probably several separators lie over the top of the emerging flux region, joining the two null clusters, as plots of fieldline connectivity show that the boundary between the four regions of magnetic connectivity lies there. We will study the formation and behaviour of the separators, and any magnetic reconnection taking place along them, in future work.

\subsection{Relationship between nulls and magnetic reconnection sites}

\begin{figure}
  \centering
  \parbox[t][0.35\textwidth][t]{\textwidth}{
    \raisebox{0.3\textwidth}{(a)}
    \includegraphics[width=0.45\textwidth,clip=true]{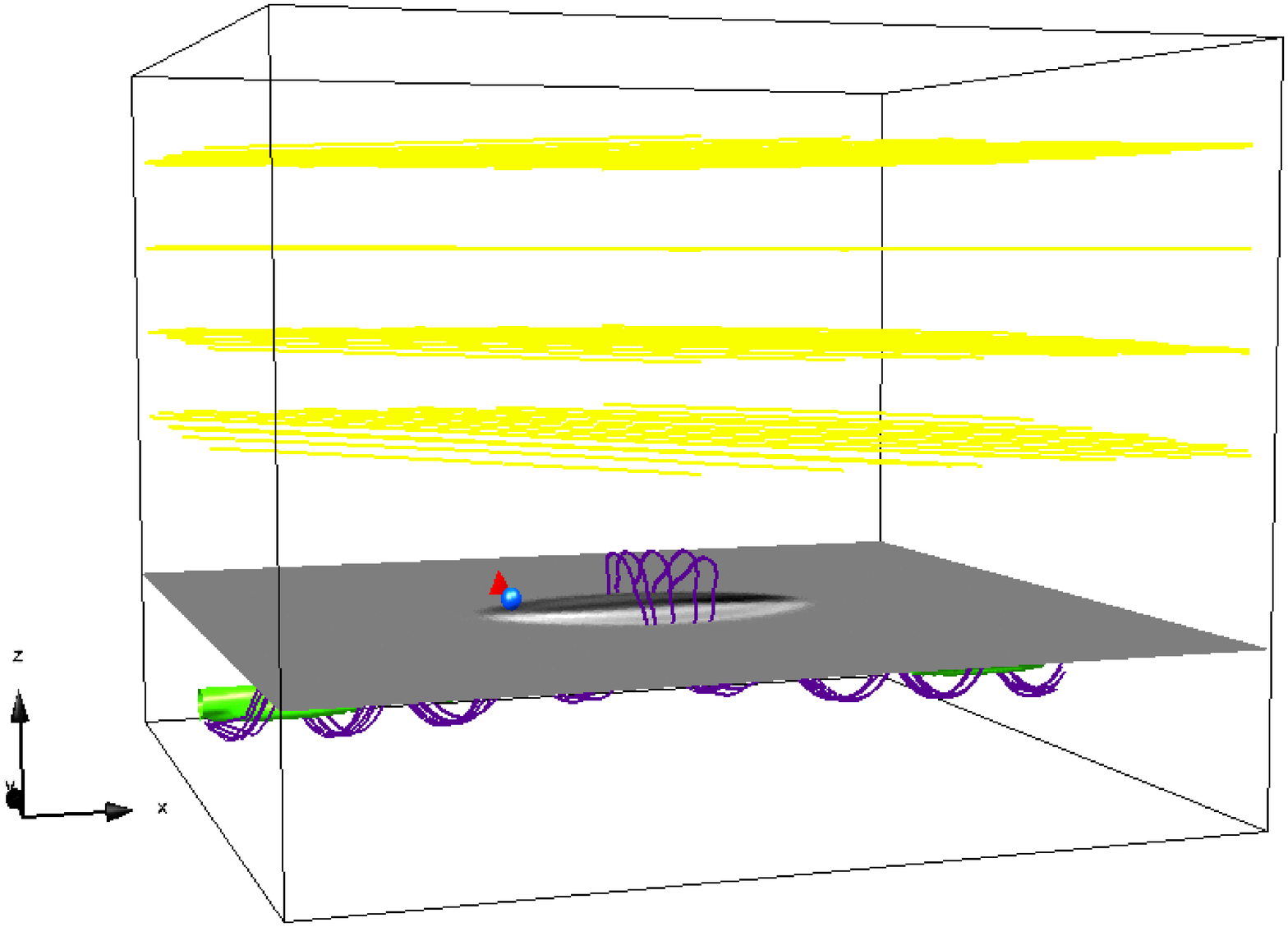}
    \raisebox{0.3\textwidth}{(b)}
    \includegraphics[width=0.45\textwidth,clip=true]{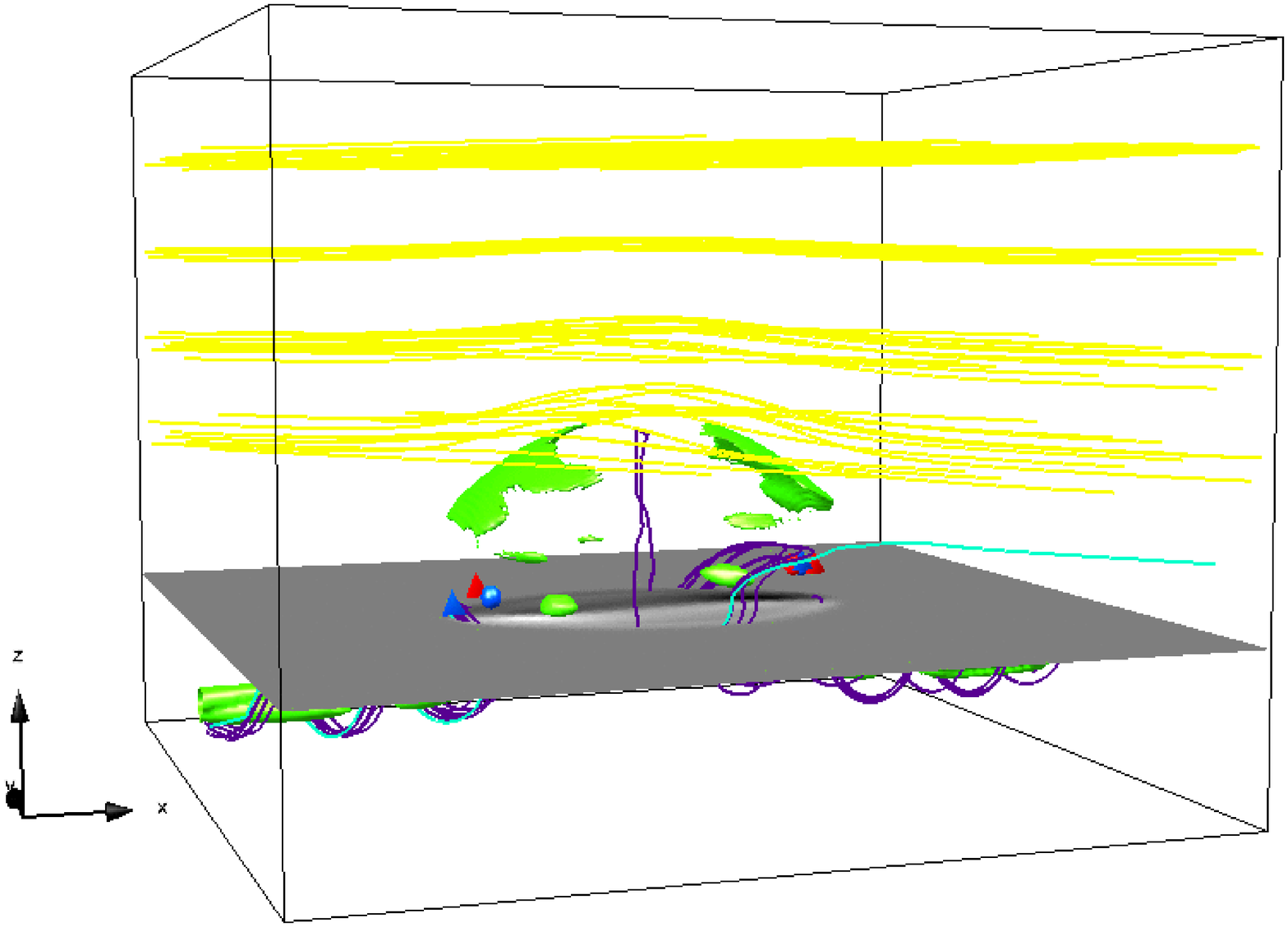}
  }
  \parbox[t][0.35\textwidth][t]{\textwidth}{
    \raisebox{0.3\textwidth}{(c)}
    \includegraphics[width=0.45\textwidth,clip=true]{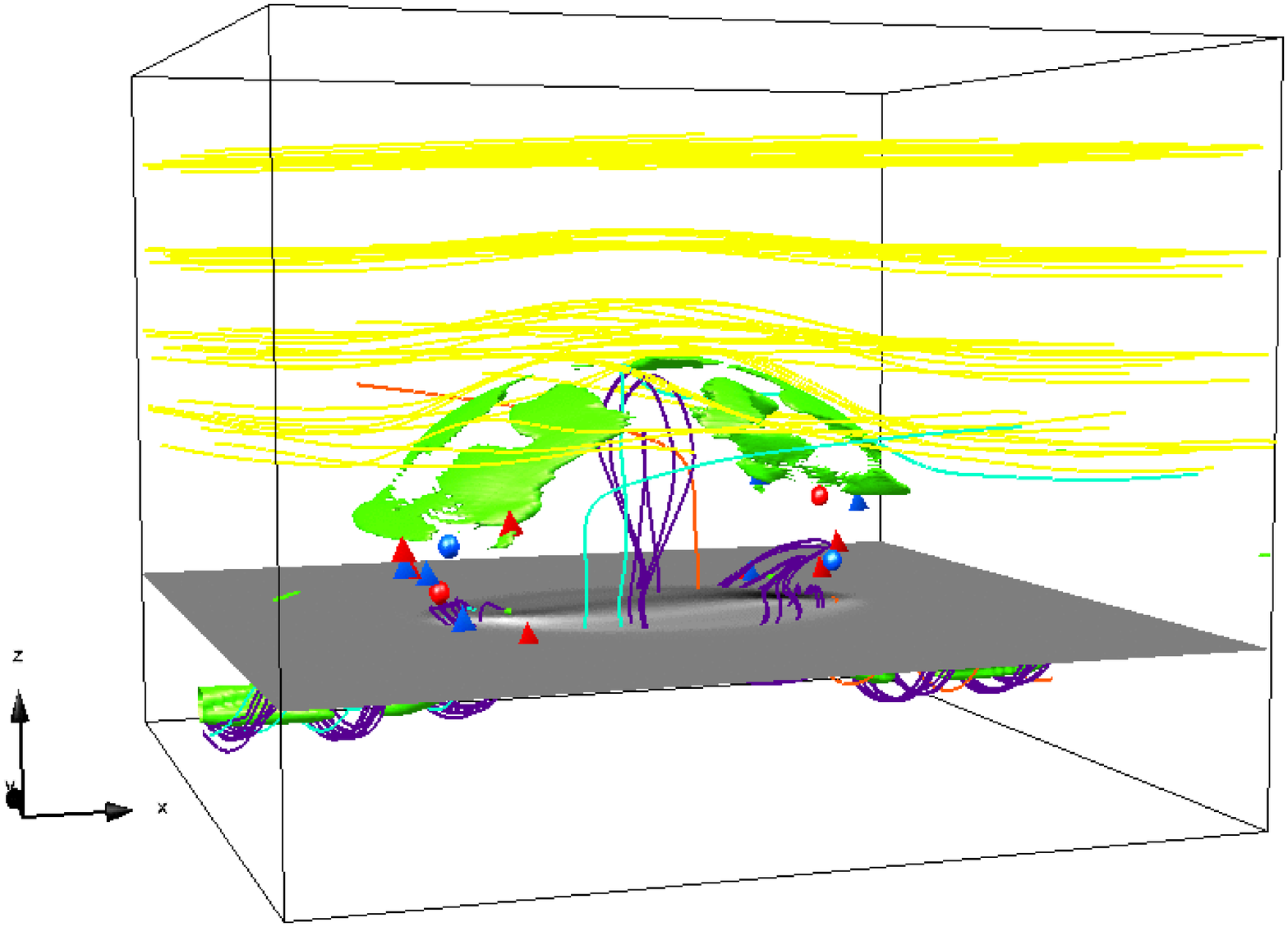}
    \raisebox{0.3\textwidth}{(d)}
    \includegraphics[width=0.45\textwidth,clip=true]{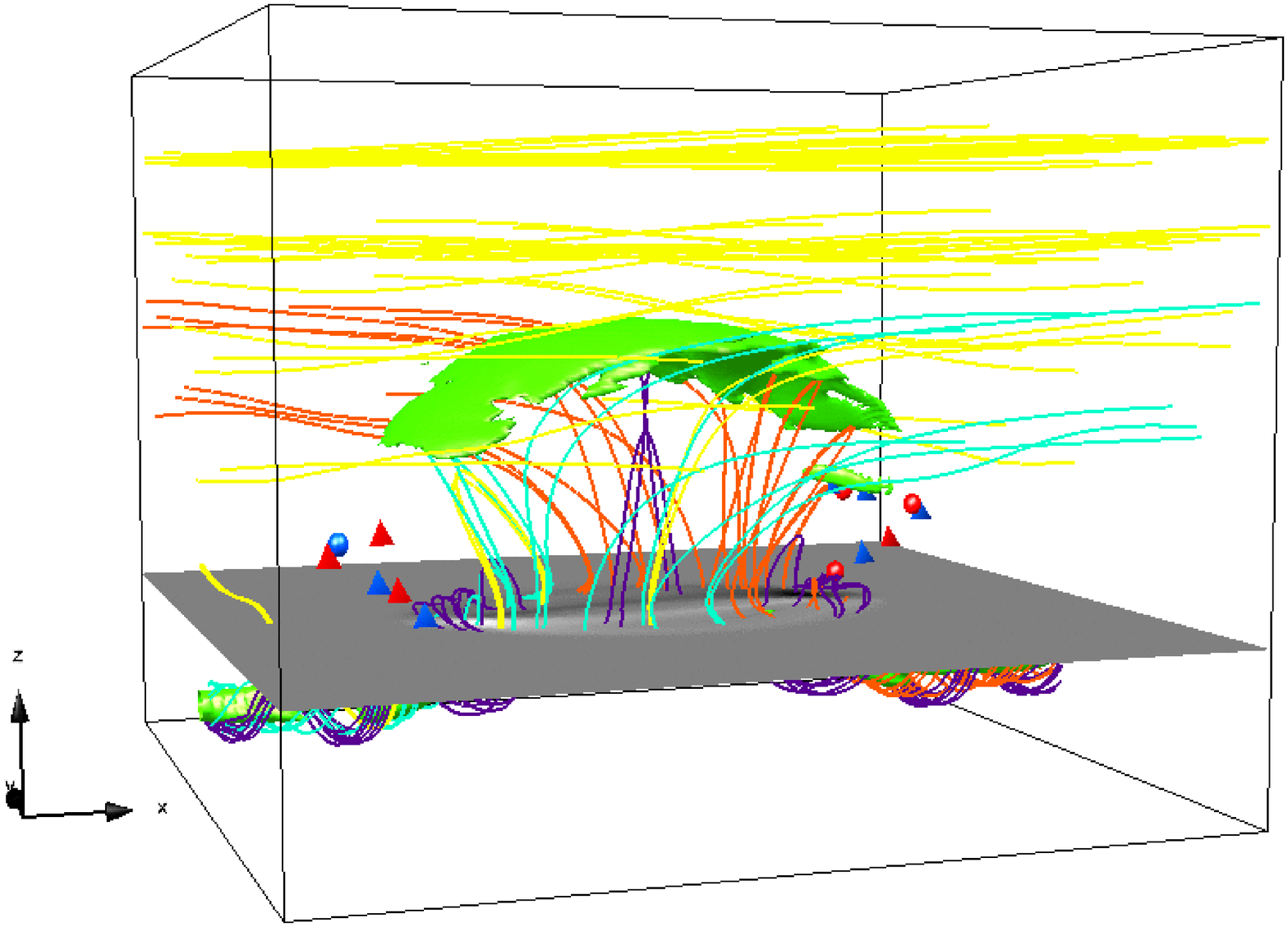}
  }
  \parbox[t][0.35\textwidth][t]{\textwidth}{
    \raisebox{0.3\textwidth}{(e)}
    \includegraphics[width=0.45\textwidth,clip=true]{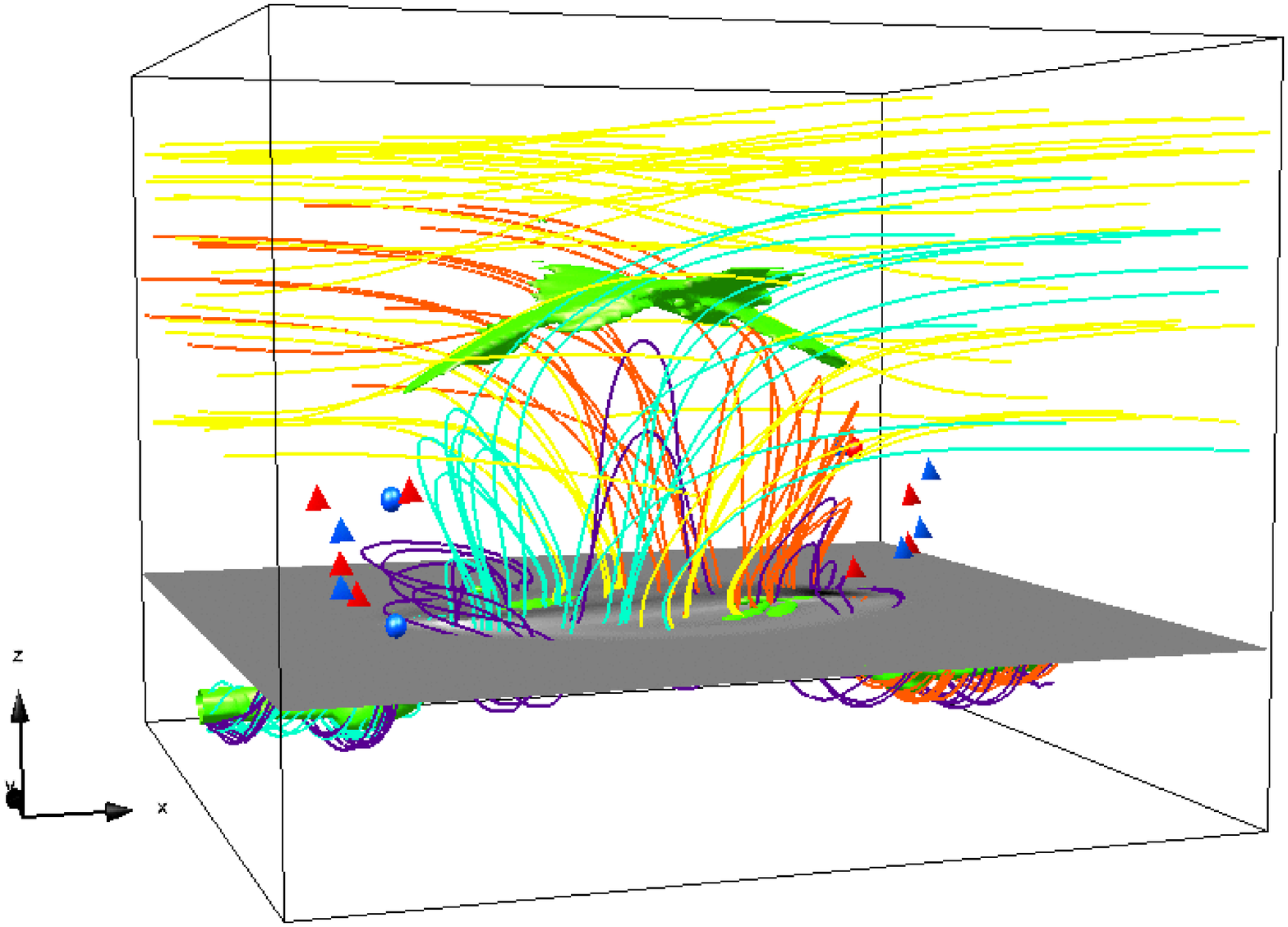}
    \raisebox{0.3\textwidth}{(f)}
    \includegraphics[width=0.45\textwidth,clip=true]{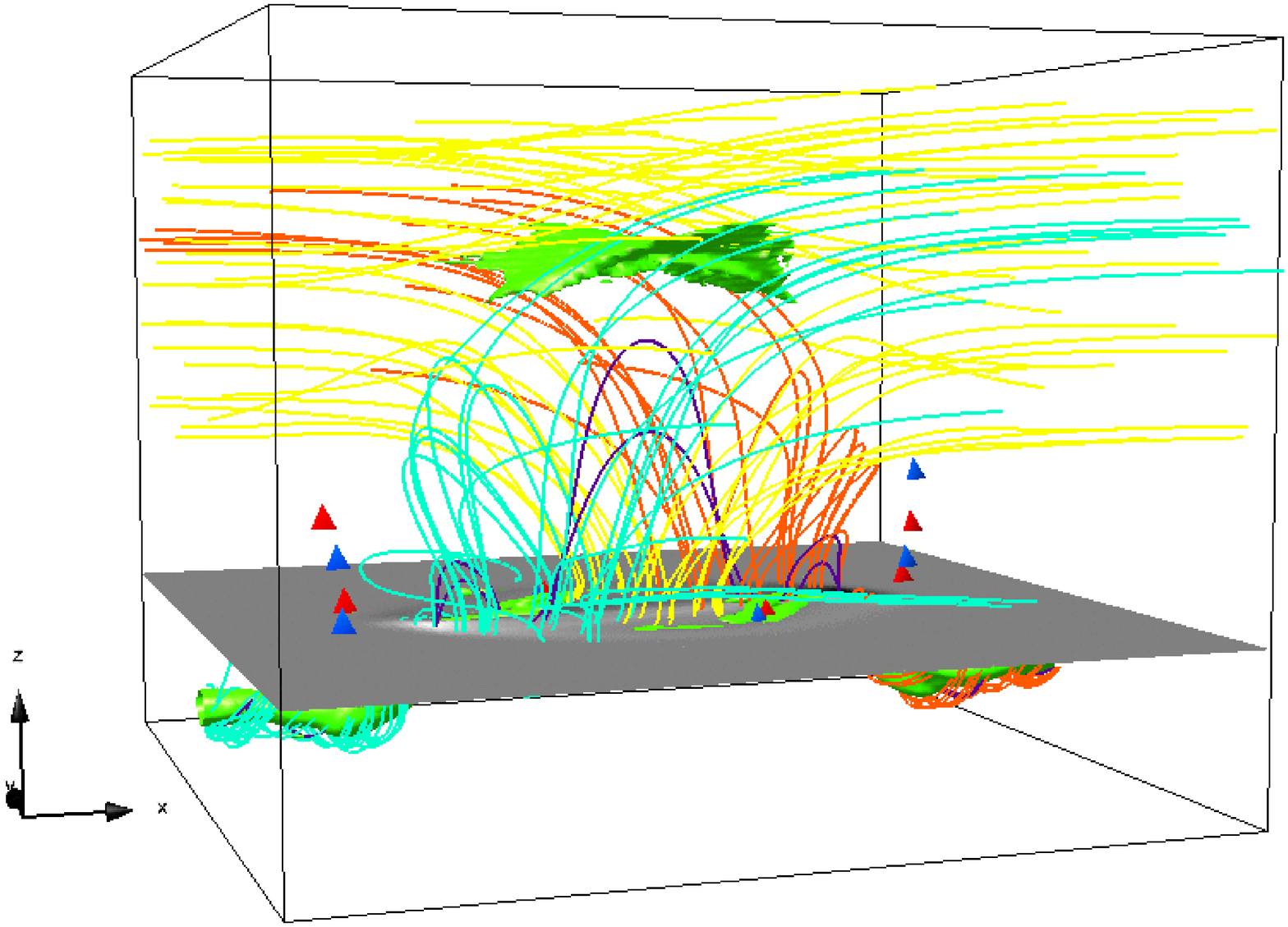}
  }
  \parbox[t][0.35\textwidth][t]{\textwidth}{
    \raisebox{0.3\textwidth}{(g)}
    \includegraphics[width=0.45\textwidth,clip=true]{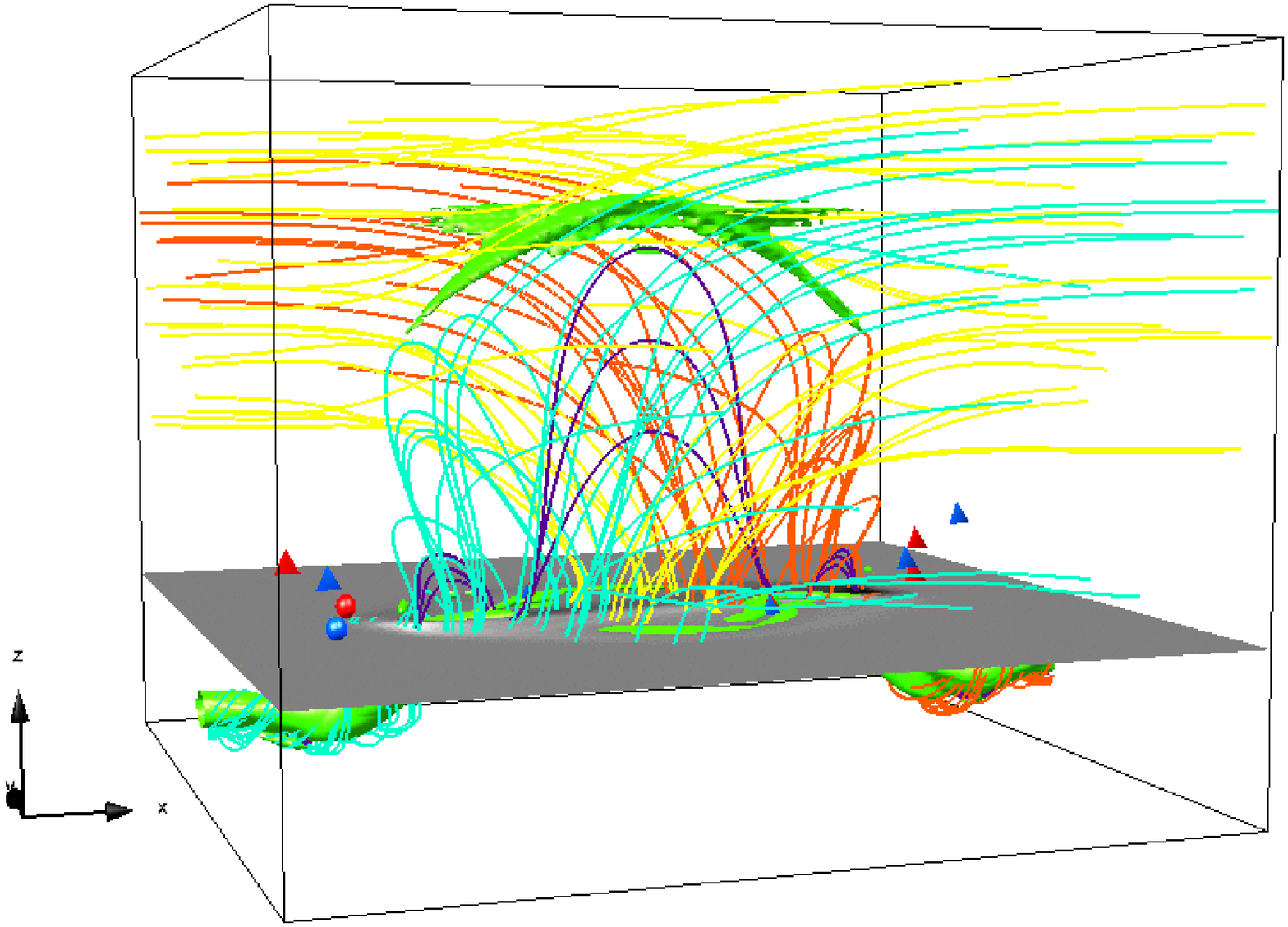}
    \raisebox{0.3\textwidth}{(h)}
    \includegraphics[width=0.45\textwidth,clip=true]{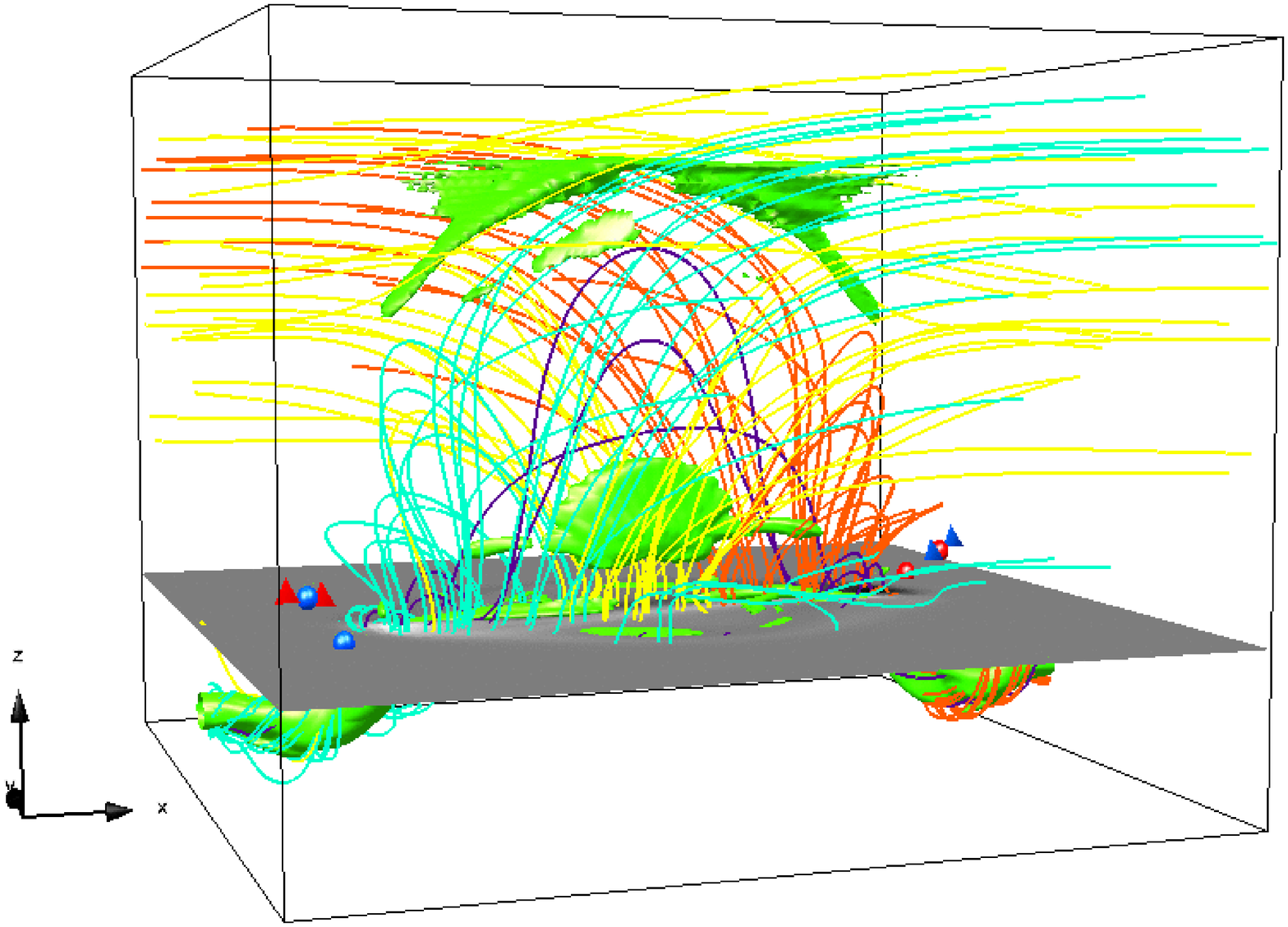}
  }
  \caption{3D views of the model as it evolves, with snapshots taken at 20, 25, 30, 35, 40, 45, 50, 55, and 60 minutes. An isosurface of parallel electric field is shown in green. The base of the photosphere is shown in the middle of the box as a (mostly grey) contour plot of vertical magnetic field (see Figure~\protect\ref{fig:b_at_photo}). Magnetic null points are displayed as tetrahedra for improper nulls and spheres for spiral nulls. Red colouring indicates a positive null and blue colouring a negative null. Finally, fieldlines are plotted in four colours: yellow for overlying field, purple for field in the flux tube, and cyan and orange for the two different connectivities of reconnected field joining the flux tube to the overlying field.}
  \label{fig:isoviz}
\end{figure}

3D snapshots of the magnetic field in the model as it evolves in time are shown in Figure~\ref{fig:isoviz}. The locations of the null clusters can be seen in relation to the flux tube, overlying, and reconnected fieldlines, which are all colour-coded as described in the figure caption. As the flux tube rises in time, the leading edge of the rising purple flux tube fieldlines moves higher up the box, and more cyan and orange reconnected fieldlines appear as the reconnection proceeds.

The figure also includes an isosurface of parallel electric field ($E_{\parallel}$). In the convection zone, the flux tube always has a high $E_{\parallel}$ simply because of the strong twist of the flux tube. In the atmosphere, as the flux tube emerges and reconnects with the overlying field, not only do null points form but also the $E_{\parallel}$ forms a cap-like sheet at the upper leading edge of the fluxtube fieldlines. This appears to be the main reconnection site in the atmosphere. Later on, an extended low-lying $E_{\parallel}$ structure also forms, straddling the two magnetic polarities in the photosphere.

\begin{figure}
  \centering
%  (a)\includegraphics[width=0.45\textwidth]{justj.eps}\hfill(b)\includegraphics[width=0.45\textwidth]{jcrossb.eps}
  \includegraphics[width=0.6\textwidth]{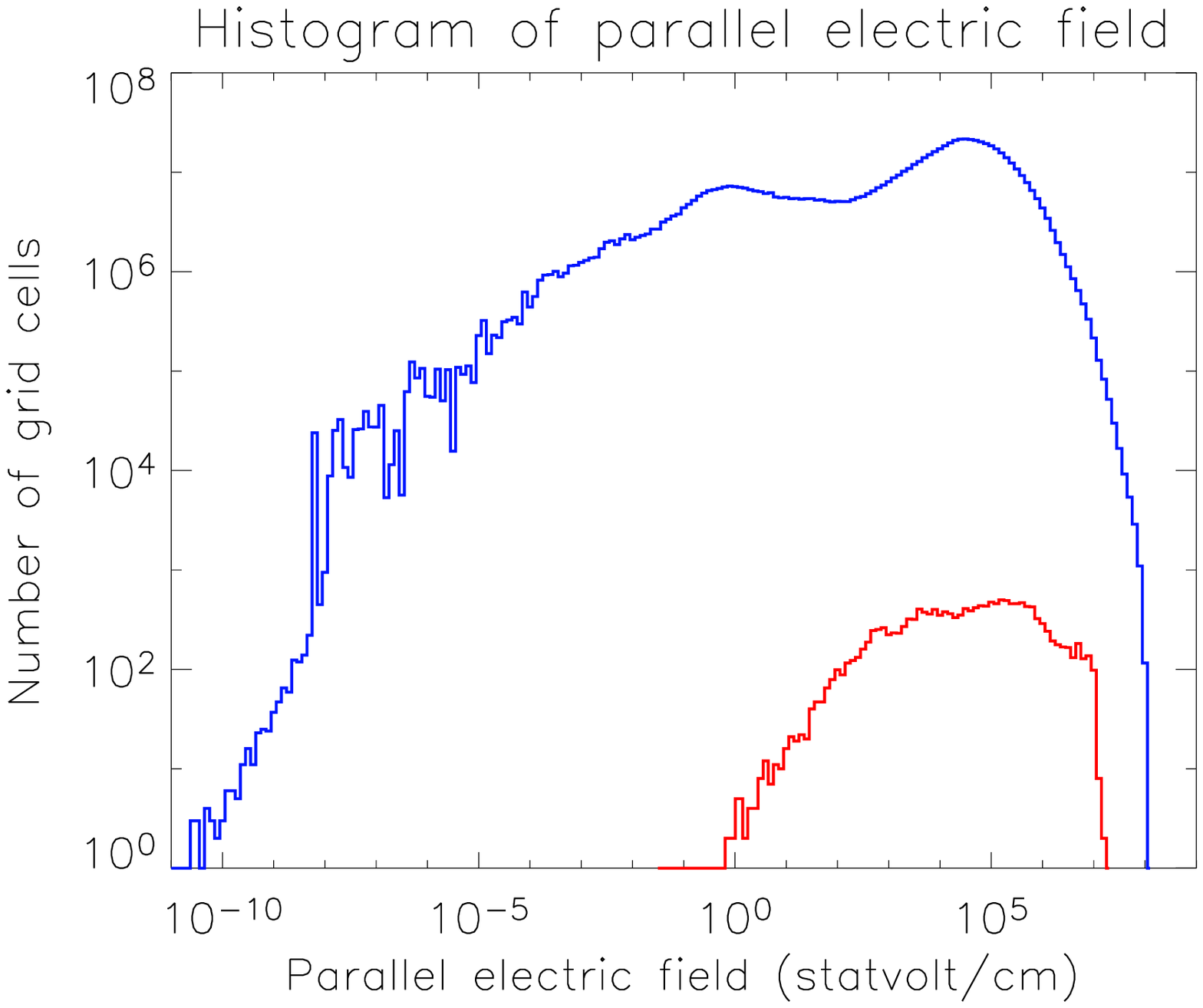}
  \caption{Histogram of the strength of the parallel electric field, everywhere in the atmosphere (blue line) and at the null points (red line). The axis scaling is logarithmic due to the wide range of values.}
  \label{fig:eparhist}
\end{figure}

Figure~\ref{fig:eparhist} shows histograms of the parallel electric field strength in the whole atmosphere (blue plot), and at grid cells containing nulls (red plot). The axis scaling on the histogram is logarithmic to allow the wide range of values for $E_{\parallel}$ to be properly visible. The maximum $E_{\parallel}$ in a null point grid cell is 14.1\% of the maximum $E_{\parallel}$ in the atmosphere. In general, the parallel electric fields near the nulls are weak compared to the strongest parallel electric fields elsewhere in the atmosphere, and they do not even register on the isosurface plot of strong $E_{\parallel}$.

A high electric current density is also required for magnetic reconnection, so we checked the strength of the current densities near the nulls, and whether the nulls are located at local maxima of current density. Not surprisingly, the results were very similar to the results for $E_{\parallel}$. The maximum current density in a null point grid cell is 19.9\% of the maximum current density in the atmosphere. We compared the maximum current density at the eight corners of the grid cell containing each null with the maximum current density at the 56 nearest-neighbouring grid cells, to check whether the local maximum current density lay at the null. In every single case, the current density was stronger further away from the null. We also repeated the calculation for the mean current density in those grid cells, and found that the mean current density increased away from the null in 82.3\% of all cases. This means that the locations of the nulls are not local maxima of current density, and so again confirms that null-point reconnection is not significant for this flux emergence experiment.

The ``cap'' of $E_{\parallel}$ between the emerged and overlying magnetic fields is obviously associated with magnetic reconnection, but it lies far away from the null points. However, since it lies along the general surface produced by the fieldlines that run close to the null clusters, we speculate that the reconnection taking place in this model may be separator reconnection \citep{1990ApJ...350..672L,1996PhPl....3.2885L,2007RSPSA.463.1097H,phg09} rather than null-point reconnection \citep{2004GApFD..98..407P,2005GApFD..99...77P}. Furthermore, the current-sheet cap lies at the boundary between the four different magnetic connectivities, where either one or more separators or possibly one or more quasi-separators must exist. Also, the weak signature of reconnection at the nulls found in this work is consistent with reconnection along separators linking the two null clusters \citep{phg09}. Alternatively, it is possible that some of the reconnection could be taking place in a ``flipping layer'' between the flux tube and the overlying field, as speculated by \citet{2005ApJ...635.1299A}. In a future work, we will investigate the importance and relative roles of both separator reconnection and non-null reconnection \citep[e.g.\ the slip-running reconnection of][]{2006SoPh..238..347A} in our flux emergence model.

\section{Discussion and conclusions}
\label{sec:conclusions}

In this paper, we have analysed the properties of the magnetic null points present in a model solar flux emergence event, and considered the implications for magnetic reconnection.

The flux emergence model that we used was first described by \citet{2007ApJ...666..516G}. In it, a twisted magnetic flux tube rises through a stratified solar convection zone and atmosphere to interact and reconnect with a horizontal overlying magnetic field. We detected a large number of magnetic null points within the resulting model magnetic field. Nulls begin to appear when the magnetic reconnection starts, and persist until the end of the entire model run. Up to 26 nulls are present at any one time.

The nulls were tracked through all the time-frames of the model, and we found that they number 305 in total. Our surprise at this unexpectedly-large number of nulls led us to speculate that some of them might simply be numerical artefacts. However, there is a lot of evidence to show that the vast majority of nulls are real, are not due to numerical artefacts, and that we can reliably identify and exclude those that are. The nulls were also classified by sign (positive or negative), and we found that the balance of signs is always in agreement with a conservation law derived from the 3D Euler equation (equation~\ref{eq:euler}), once the spurious nulls are excluded. The lifetimes of the nulls were calculated, and the majority of the nulls were found to persist for many multiples of the mean timestep between frames. Finally, the stability of the nulls was also assessed by calculating how fast their spine eigenvectors were rotating. The vast majority of nulls were found to have remarkably steady spine directions, meaning that they are very stable.

The nulls are located throughout the model photosphere and transition region. They never reach as high as the bottom of the model corona. The reason for this is because in the vast majority of our model corona the magnetic field is very simple (an overlying horizontal magnetic field) and the emergence of new flux only creates complexity in small localised regions low down in the atmosphere. \citet{2009SoPh..254...51L} have investigated the number of nulls that occur in the solar atmosphere and showed that indeed their number falls off quickly the further you get from the photosphere, as the complexity of the magnetic field decreases with height. The nulls in our experiment form just after the reconnection has started, in two clumps low down at the boundary between the emerging and the overlying magnetic flux. As the emerging region expands, the nulls are pushed out towards the side boundaries of the box, but they never reach or cross these boundaries during the model run.

A parallel electric field is a necessary requirement for reconnection, so we investigated its nature in the vicinity of the nulls. Regions of strong parallel electric field are found in the model at the peak of the emerging fieldlines, and also (later on) lower down in the atmosphere. Although many of the nulls have some associated parallel electric field, we found that this is weak compared to the strongest parallel electric fields elsewhere in the atmosphere, which are concentrated far from the nulls in the regions just described. Null-point reconnection is therefore ruled out as the predominant type of magnetic reconnection taking place during the emergence event. Clearly though, the reconnection is associated with either separators from these nulls or reconnection in the absence of nulls. We plan to investigate these possibilities more thoroughly in future.
% Our present hypothesis is that the reconnection is taking place along one or more separators joining the two null clusters. The reason for this is that the regions of high parallel electric field area located at the top of the rising flux tube, where fieldlines of all four different magnetic connectivities are present, and so one or more separators (which must bound four regions of differing magnetic connectivity) should be present there. Separator reconnection has been found to be the dominant type of reconnection in other, simpler, models of magnetic field evolution \citep{2007RSPSA.463.1097H,phg09}, and in future work we will go on to determine its importance for emerging flux.

\begin{acks}
The authors would like to thank Dr.\ D.\ Pontin for his insightful comments, and also the anonymous referee for his thoughtful suggestions which helped us to improve the paper. Computational time on the UKMHD Linux clusters in St Andrews (STFC and SRIF funded) is gratefully acknowledged. This work was supported by the European Commission through the SOLAIRE Network. RCM is financially supported by the University of St Andrews Solar Group STFC Rolling Grant.

\end{acks}

\bibliographystyle{spr-mp-sola-cnd}
\setlength{\bibsep}{0pt}
\bibliography{maclean}

\end{article}

\end{document}